\documentclass{aa}  

\usepackage{graphicx}
\usepackage{txfonts}
\usepackage{subcaption}
\usepackage[colorlinks=true,citecolor=blue]{hyperref}
\usepackage{xcolor}
\usepackage{ulem}
\newcommand{\kms}{km~s$^{-1}$}
\newcommand{\form}{\mbox{H$_{2}$CO}~}
\newcommand{\meth}{\mbox{CH$_{3}$OH}~}
\newcommand{\degree}{$^\circ$}
\DeclareRobustCommand{\ion}[2]{\textup{#1\,\textsc{\lowercase{#2}}}}
\DeclareRobustCommand{\kms}{km\,${\rm s}^{-1}$}
\DeclareRobustCommand{\jyb}{Jy\,beam${}^{-1}$\,}
\newcommand{\hii}{\ion{H}{ii}}

\begin{document}

   \title{A global view on star formation: The GLOSTAR Galactic plane survey}

   \subtitle{I. Overview and first results for the Galactic longitude range $28^\circ<l<36^\circ$}

   \author{A. Brunthaler \inst{1},
          K. M. Menten \inst{1},
          S. A. Dzib \inst{1},
          W. D. Cotton \inst{2,3},
          F. Wyrowski \inst{1},
          R. Dokara \inst{1}\thanks{Member of the International Max Planck Research School (IMPRS) for Astronomy and Astrophysics at the Universities of Bonn and Cologne},
          Y. Gong \inst{1},
          S-N. X. Medina \inst{1},
          P. M{\"u}ller \inst{1},
          H. Nguyen \inst{1,\star},
          G. N. Ortiz-Le{\'o}n \inst{1}, 
          W. Reich \inst{1},
          M. R. Rugel \inst{1},
          J.\,S.\,Urquhart \inst{4},
          B. Winkel \inst{1},
          A. Y. Yang \inst{1},
          H. Beuther \inst{5},
          S.\,Billington \inst{4},
          C. Carrasco-Gonzalez \inst{6},
          T. Csengeri \inst{7},
          C. Murugeshan \inst{8},
          J. D. Pandian \inst{9},
          N. Roy \inst{10}}

   \institute{Max-Planck-Institut f\"ur Radioastronomie, 
              Auf dem H\"ugel 69, 53121 Bonn, Germany\\
              \email{brunthal@mpifr-bonn.mpg.de}
    \and
    National Radio Astronomy Observatory, 520 Edgemont Road, Charlottesville, VA 22903, USA
    \and
           South African Radio Astronomy Observatory, 2 Fir St, Black River Park, Observatory 7925, South Africa
    \and
    Centre for Astrophysics and Planetary Science, University of Kent, Ingram Building, Canterbury, Kent CT2 7NH, UK
    \and
    Max Planck Institute for Astronomy, Koenigstuhl 17, 69117 Heidelberg, Germany
    \and
    Instituto de Radioastronomía y Astrofísica (IRyA), Universidad Nacional Autónoma de México Morelia, 58089, Mexico
    \and {Laboratoire d'astrophysique de Bordeaux, Univ. Bordeaux, CNRS, B18N, allée Geoffroy Saint-Hilaire, 33615 Pessac, France} 
    \and
    Centre for Astrophysics and Supercomputing, Swinburne University of Technology, Hawthorn, Victoria 3122, Australia
    \and
    Department of Earth \& Space Sciences, Indian Institute of Space Science and Technology, Trivandrum 695547, India
    \and
    Department of Physics, Indian Institute of Science, Bengaluru 560012, India
    }
  \authorrunning{Brunthaler et al.}

   \date{}

 
  \abstract
   {}
   {Surveys of the Milky Way at various wavelengths have changed our view of star formation in our Galaxy considerably in recent years. In this paper we give an overview of the GLOSTAR survey, a new survey covering large parts (145 square degrees) of the northern Galactic plane using the Karl G. Jansky Very Large Array (JVLA) in the frequency range  $4-8$ GHz and the Effelsberg 100-m telescope. This provides for the first time a radio survey covering all angular scales down to 1.5 arcsecond, similar to complementary near-IR and mid-IR galactic plane surveys. We outline the main goals of the survey and give a detailed description of the observations and the data reduction strategy.}
   {In our observations we covered the radio continuum in full polarization, as well as the 6.7 GHz methanol maser line, the 4.8~GHz formaldehyde line, and seven radio recombination lines. The observations were conducted in the most compact D configuration of the VLA and in the more extended B configuration. This yielded spatial resolutions of 18$"$ and 1.5$"$ for the two configurations, respectively. We also combined the D configuration images with the Effelsberg 100-m data to provide zero spacing information, and we jointly imaged the D- and B-configuration data for optimal sensitivity of the intermediate spatial ranges.}
   {Here we show selected results for the first part of the survey, covering the range of 28$^\circ<l<36^\circ$ and $|b|< 1^\circ$, including the full low-resolution continuum image, examples of high-resolution images of selected sources, and the first results from the spectral line data.
}
   {}

   \keywords{surveys, ISM: general, (ISM:) HII regions, ISM: supernova remnants, radio lines: ISM, radio continuum: general
               }

   \maketitle
%

\section{Introduction}
Stars with more than about ten solar masses dominate galactic ecosystems. Understanding the circumstances
of their formation is one of the great challenges of modern astronomy.  In recent years, our view of
massive star-forming regions has been dramatically  changed by Galactic plane surveys covering  the infrared (e.g.,\ Galactic Legacy Infrared Mid-Plane Survey Extraordinaire (GLIMPSE; \citealp{churchwell2009}), MIPSGAL (\citealp{carey2009}), Hi-GAL (\citealp{Molinari2010})); (sub-)millimeter (e.g., APEX Telescope Large Area Survey of the Galaxy (ATLASGAL; \citealp{SchullerMenten:2009aa,csengeri2014}), Bolocam Galactic Plane Survey (BGPS; \citealp{ref-bgps}), JCMT Plane Survey (JPS; \citealp{moore2015,Eden2017}), the Millimetre Astronomy Legacy Team 90\,GHz  (MALT90; \citealt{jackson2013})); and radio wavelength ranges  (e.g., the Multi-Array Galactic Plane Imaging Survey (MAGPIS; \citealp{Becker1990, becker1994}), the Sino-German 6 cm survey \citep{Han2015}\footnote{Data from this and various other large-scale centimeter-wavelength surveys can be accessed via the MPIfR  Survey Sampler at \url{https://www.mpifr-bonn.mpg.de/survey.html}}, the Coordinated Radio and Infrared Survey for High-Mass Star Formation (CORNISH; \citealt{HoarePurcellChurchwell2012, PurcellHoareCotton2013}), The \ion{H}{i}, OH and Radio Recombination line survey of the Milky Way (THOR; \citealt{thor2016,Wang2020}), the Southern Galactic Plane Survey \citep{McClure2005}, the VLA Galactic Plane Survey (VGPS; \citealt{Stil2006}), the Canadian Galactic Plane Survey \citep{Taylor2003}, the H$_2$O Southern Galactic Plane Survey (HOPS; \citealt{hops2011}), the Methanol Multibeam Survey (MMB; \citealt{GreenCaswellFuller2009}), the Galactic Ring Survey (GRS; \citealt{Jackson_2006})). These surveys allow us for the first time to study {all} the evolutionary stages of massive star formation (MSF) in an unbiased way (e.g., \citealt{koenig2017, elia2017, UrquhartKonig:2018aa}).

\begin{table*}
\centering
\small
\begin{tabular}{ccccccc}
Line & Frequency &  Bandwidth & Chan.& Resolution & Coverage & 1$\sigma$ 
rms in 15 sec\\
     & [MHz]     &    [MHz]   &  \& pol. prod. & [\kms]&[\kms] & [mJy beam$^{-1}$]\\
\hline
\hline
Continuum&4200--5200 & 8$\times$128  &8$\times$64$\times$4&-&-&0.09 \\
H114$\alpha$ & 4380.954 & 8 &  128$\times$2 & 4.3 & 547 & 11\\
H113$\alpha$ & 4497.776 & 8 &  128$\times$2 & 4.2 & 533 & 11\\
H112$\alpha$ & 4618.789 & 8 &  128$\times$2 & 4.1 & 529 & 11\\
H$_2$CO      & 4829.660 & 4 & 1024$\times$2& 0.24 & 248 & 45\\
H110$\alpha$ & 4874.157 & 8 &  128$\times$2 & 3.8 & 492 & 11 \\
\hline
Continuum&6400--7400 & 8$\times$128  &8$\times$64$\times$4&-&-&0.08\\
CH$_3$OH    & 6668.518 & 8 & 2048$\times$2& 0.18 &  360 & 38\\
H99$\alpha$  & 6676.076 & 8 &  128$\times$2& 2.8 & 359 & 11\\
H98$\alpha$  & 6881.486 & 8 &  128$\times$2& 2.7 & 348 & 11\\
H96$\alpha$  & 7318.296 & 8 &  128$\times$2 & 2.5 & 328 & 11\\
\hline
\hline
\end{tabular}
\caption{List of observed lines, bandwidth, number of channels and polarization products, channel spacings, velocity coverage, and theoretical sensitivity for two visits of one pointing (for the full 1 GHz for the continuum and 1 channel for spectral lines) of the VLA observations. For $l$=58\degree--60\degree\ the RRLs H115$\alpha$ (4268.142 MHz), H102$\alpha$ (6106.855 MHz), and H103$\alpha$ (5931.544 MHz) were observed instead of H110$\alpha$ and H96$\alpha$. The central LSR velocity changed for different parts of the survey based on longitude-velocity plots of CO in the Milky Way.}
\label{correl}
\end{table*}
With the exciting results of the submillimeter and far-infrared (FIR) surveys from the ground (ATLASGAL) and space (Hi-GAL), the massive and cold dust clumps from which massive clusters form are now being  detected galaxy-wide. Complementary to these surveys, the centimeter studies using the VLA allow very powerful and comprehensive radio-wavelength surveys of  the ionized and the molecular tracers of star formation in the Galactic plane. There have been a number of VLA surveys of  the inner part of the first quadrant of the Galactic plane starting with the pioneering 20\,cm (\citealt{Becker1990}) and 6\,cm surveys (\citealt{becker1994}) that are now combined in the MAGPIS project.\footnote{https://third.ucllnl.org/gps/} More recently the CORNISH team have mapped the northern GLIMPSE region at 5\,GHz with  higher resolution and sensitivity than MAGPIS. CORNISH was tailored to specifically  focus on identifying and  parameterizing the ultra-compact (UC) \hii\ region stage of MSF (e.g., \citealt{PurcellHoareCotton2013, kalcheva2018}), although it has also been used to investigate the Galactic population of planetary nebulae (PNe; \citealt{irabor2018}). These studies have identified many thousands of radio sources located towards the Galactic plane, and follow-up work has allowed many hundreds of Galactic radio sources to be identified and studied (e.g., \citealt{urquhart2013_cornish, cesaroni2015,kalcheva2018, irabor2018}). However, these interferometric snapshot surveys have been limited by the historically small bandwidths available ($\sim$50\,MHz) and sensitivity to emission on larger angular scales. These surveys have therefore been restricted to continuum studies of relatively compact structures (i.e., $<$20\arcsec\ for CORNISH) and are therefore unable to provide us with a global view of star formation in the Galaxy.
 
The scope of radio-wavelength Galactic plane surveys and of radio interferometry in general, has been vastly increased with the expansion of the VLA into the Karl G. Jansky Very Large Array, (JVLA\footnote{Initially known as the  Expanded VLA (EVLA)}). The VLA expansion project, beginning in 2001, with Early Science programs starting in early 2010, replaced the  1970s electronic and digital hardware with state-of-the art technology \citep[for a description, see][]{Perley2011}. The expansion provided, among other things, continuous frequency coverage from 1 to 50 GHz with receivers with much wider bandwidth and better noise performance \citep[see Table 1 of][]{Perley2011}. With the Wideband Interferometric Digital ARchitecture (WIDAR) correlator several metrics of the array's spectroscopy performance improved by orders of magnitude in terms of instantaneous frequency coverage and spectral resolution \citep[see Table 2 of][]{Perley2011}. Finally, of key importance for large area surveys, the ``dead time'' between any two observations (e.g., with a change of position) was much reduced from the 20 s  typical of the old VLA down to the slew and settling time of the antennas, allowing efficient mosaicking strategies. All this makes the new VLA an  extremely sensitive and efficient survey instrument. This is especially true for projects that combine continuum and spectral line observations, which was realized early on by \cite{Menten2007}, who described a project that has essentially evolved into the GLOSTAR survey.

The GLOSTAR survey has been designed to address the limitation placed on these previous surveys using the vastly improved capabilities of the upgraded VLA. Our goal was to use the extremely wide-band (4--8 GHz) C-band receivers of the VLA for an unbiased survey to find and characterize  star-forming regions in the Galaxy. All fields are observed in the most compact D configuration of the VLA with a resolution of $\sim 18''$ to have good surface brightness sensitivity for extended structures and in the more extended B configuration that provides a ten times higher spatial resolution to investigate compact sources. The VLA data are then complemented by observations with the Effelsberg 100-m telescope to provide zero spacings for the D-configuration data. Finally, the data from the  D and B configurations are combined (D+B) to provide the best possible images of sources with intermediate sizes by combining the surface brightness sensitivity of the D configuration with the high resolution of the B configuration.

This survey of the Galactic plane detects tell-tale tracers of star formation: compact, ultra-compact,  and hyper-compact H{\sc ii} regions and molecular masers that trace different stages of early stellar evolution and pinpoints the very centers of the early phase of star-forming activity. Combined with the submillimeter and infrared surveys, and our ongoing work in the Bar and Spiral Structure Legacy (BeSSeL)
Survey\footnote{see http://bessel.vlbi-astrometry.org for more details} to measure distances by trigonometric parallaxes to most of the dominant star-forming regions in the Galaxy, it offers a nearly complete census of the number, luminosities, and masses of massive star-forming clusters in a large range of evolutionary stages and provides a unique  data set with true legacy value for a global perspective on star formation in our Galaxy. In addition to information on high-mass star formation, the GLOSTAR survey allows efficient identification and imaging of supernova remnants, planetary nebulae, and numerous extragalactic background sources.

\section{The GLOSTAR VLA survey overview}

\subsection{Survey coverage}
The full survey covers the Galactic plane  in the range  $-2^\circ<l<60^\circ$ and $|b|<1^\circ$, as well as the Cygnus X star-forming complex at a range of $76^\circ<l<83^\circ$ and $-1^\circ<b<2^\circ$, or 145 square degrees in total. We needed about 341 individual pointings to cover one square degree, or 49463 pointings in total.

To date, the full survey region has been observed in the D configuration. For the higher-spatial-resolution observations in  B configuration the region in the range $40^\circ<l<58^\circ$ has not been observed yet. For the Galactic center region between $-2^\circ<l<12^\circ$, we use the hybrid DnC and BnA configurations when possible, for which the northern arm of the array is in the more extended configuration to ensure a more circular synthesized beam. Since the hybrid configurations were not available after 2016, we use a combination of D and C configuration observations for the range  $6^\circ<l<9^\circ$, and a combination of B and A configuration observations for $l<10^\circ$ instead.

In this paper we show results and examples of the first 16 square degrees in the range  $28^\circ<l<36^\circ$ and $|b|<1^\circ$. The observations of this range were made under proposal code 13A-334 in the D and B configurations between 2013 April 6 and 2013 May 2, and 2013 September 29 and 2014 January 31, respectively.

\subsection{Correlator setup}

In GLOSTAR we observe two 1 GHz  basebands in full polarization mode to measure the C-band continuum emission. Higher-frequency-resolution  correlator windows are used to cover the  most prominent methanol (CH$_3$OH) maser emission line at 6.7~GHz (5$_{1}$--6$_{0}$ A$^{+}$), seven radio recombination lines (RRLs) to study the kinematics of the ionized gas, and the 4.829 GHz (1$_{1,0}$--1$_{1,1}$) transition of formaldehyde H$_2$CO, where absorption measurements can solve distance ambiguities. The lines are observed in dual polarization mode with channel spacings of \mbox{3.9 kHz} for H$_2$CO and CH$_3$OH, and 62.5 kHz for the RRLs. The bands are centered at a local standard of rest (LSR) velocity of 50 \kms. This setup avoids the strong persistent radio frequency interference (RFI) from a microwave link seen at 6.3 GHz at the VLA site, and allows for an estimation of the spectral index  between the lower and higher basebands at 4.7 and 6.9 GHz, respectively. Table~\ref{correl} lists the details of the  setup used, together with the expected sensitivities for two scans with a total integration time of 15 seconds for each pointing. We note that the final sensitivities are expected to be at least a factor of two better, since each field is also covered by six neighboring fields. The continuum data is recorded with 16 spectral windows with 64 channels each, resulting in a bandwidth of 2 MHz for each channel. We use a correlator dump time of 2 seconds for all configurations. The high spectral and time resolution avoids any significant time and bandwidth smearing even when imaging the full primary beam. This setup results in a total data rate of 17.3 MB/s. 

For the Galactic Center region from $-2^\circ<l<2^\circ$, we use a slightly modified correlator setup, where we double the bandwidth of the spectral line observations at the cost of a lower velocity resolution. This was necessary due to the very broad velocity range seen in the Galactic Center.

In our first pilot observations in the range $58^\circ<l<60^\circ$ we also used a different setup. Here the two continuum sub-bands covered the ranges  $4.1-5.1$ GHz and $5.9-6.9$ GHz, and we had one additional high-resolution band covering the 6.035 GHz OH transition. However, due to very strong RFI at 6.3 GHz, we decided to change the frequency setup to avoid the strongest RFI.

\begin{figure}
\includegraphics[width=9cm]{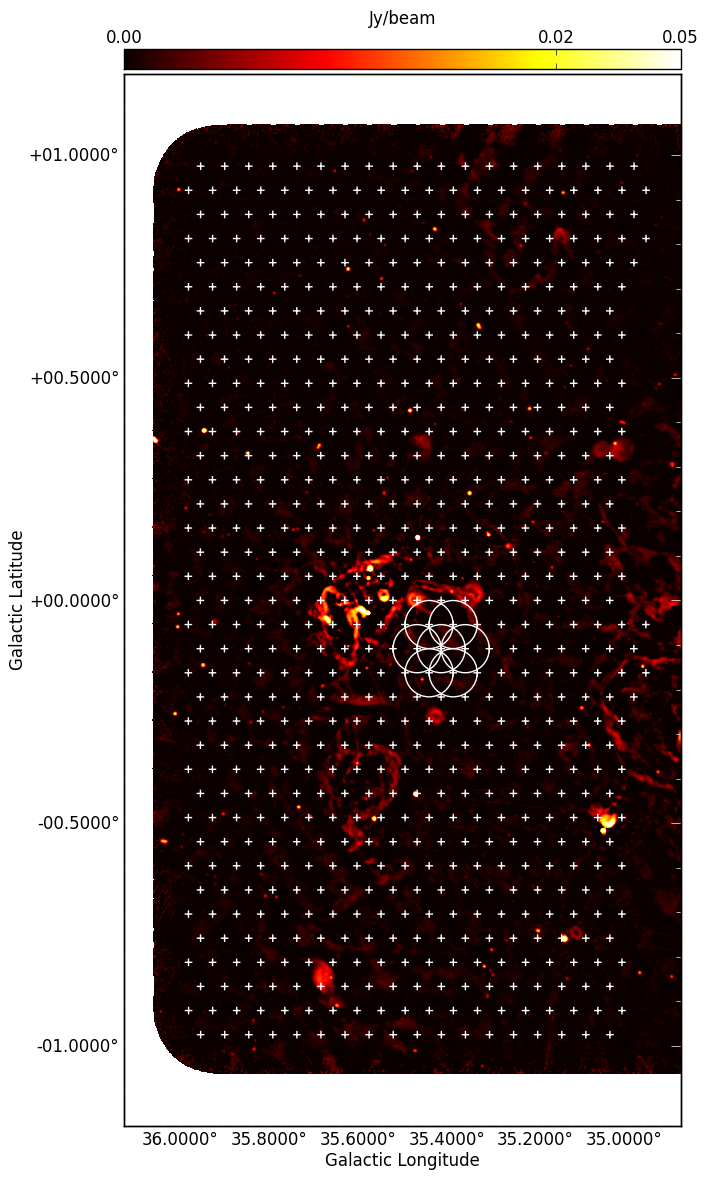}
\caption{Pointing configuration of a single observation spanning the range of $35^\circ<l<36^\circ$ and $|b|<1^\circ$. The background image shows our D-configuration continuum data. Also shown are the primary beam sizes at the center frequency of the higher frequency continuum band (6.9 GHz) of one central pointing and its six neighboring pointings.}
\label{fig:pointings}
\end{figure}

\subsection{Mosaicking strategy}

The imaging of large regions on the sky with the VLA at medium radio frequencies (4--8 GHz) in a reasonable observing time requires very short scans per pointing. However, for short scans the overhead for slewing
and settling of the antennas becomes an important factor for the total time estimate. On-the-fly mapping 
minimizes the overhead, but comes at the expense of highly increased data rates. If we move within a few
seconds over one primary beam of the antennas, we have to sample the data with a short integration time of 
typically 0.5 seconds to avoid smearing effects.  This  results in a data rate exceeding 60 MB/s, which
is beyond the limit for regular VLA observations and much higher than the data rate of 17 MB/s that we use. Furthermore,
we cannot  time-average the data easily
in post processing, which dramatically increases the required computing time. Therefore, we used
traditional pointings over on-the-fly mapping.

Since the settling time of the VLA antennas is longer if the antennas move in azimuth compared to only elevation slews, we   developed an observing scheme that minimizes the slews in azimuth. The CASA\footnote{https://casa.nrao.edu/} task {\it makeschedule}, which is now available as a contributed task\footnote{http://casaguides.nrao.edu/}, can generate schedules that minimize the slew and settling times. Using this method we were able to get, on average, 7-8 seconds of good data from an observing scan of 11 seconds (i.e., an efficiency of $\sim$70\%). In one five-hour observation, we mapped a $2^\circ\times1^\circ$ strip across the Galactic plane from $|b|<1^\circ$ with
typically 676 single pointings. Each pointing was observed twice for 11 seconds, yielding a total integration
time of $\sim$15 seconds. The pointings were on a hexagonal grid with a grid spacing of
$\theta_{hex}=$3.25$^\prime$ (see Fig.~\ref{fig:pointings} for an example). This corresponds to $\theta_B/2$,
where $\theta_B=6.5'$ is the primary beam at the central
frequency of the higher continuum band (6.9 GHz).  With this setup we have better than
Nyquist sampling for most of the data (Cornwell 1988), and also a better sampling than the  $\theta_B/\sqrt2$
used in other surveys, such as the National Radio Astronomy Observatory (NRAO) VLA Sky Survey
(NVSS)~\citep{CondonCottonGreisen1998} and
CORNISH~\citep{HoarePurcellChurchwell2012}. Having a denser grid has the advantage that the six neighboring pointings will contribute to the final image. Since neighboring rows are observed at different times, this will provide much better uv-coverage than a single pointing. The improved uv-coverage will be important for fields with strong and extended emission.

We used 3C\,286 (and sometimes 3C\,48 as a backup) as flux density and bandpass calibrator. A phase calibrator was
observed every 15-20 minutes in D configuration and every 5-10 minutes in B configuration.

\section{VLA data calibration}
\subsection{Continuum data}

The continuum data were calibrated with the Obit package~\citep{Cotton2008} using instrumental calibration signals and observations of celestial sources. The primary flux density and polarization calibrator was 3C\,286 using the flux density scale of~\cite{PerleyButler2013a} and the
polarization of \cite{PerleyButler2013b}. A secondary calibrator was chosen near the region being observed. Gain and delay calibrations were kept in a table and a single bandpass table was used. Flagging consisted of a list of data to be excluded and kept in a table. Calibration and editing consisted of the following steps: 
\begin{enumerate}
\item{\bf Convert from Archive format.}
The data were converted from archive format (ASDM/BDF) to AIPS format.
\item{\bf Initial flagging.}
The flag table was initialized to the flags determined by the online system. Data were compared to running medians in time (each scan is at least three or four times the correlator dump time of 2 seconds) and frequency flagging the outliers. 
Antennas were flagged when shadowed by other antennas.
\item{\bf Switched power calibration.}
An amplitude correction based on the switched power signal corrected for variations in system temperature.
\item{\bf Parallactic angle correction.}
A time-dependent correction was applied to the calibration table for the parallactic angle of the antennas.
\item{\bf Delay calibration.}
Group delay corrections were determined from calibrator data and applied to all sources.
\item{\bf Bandpass calibration.}
Bandpass corrections were determined from the primary flux density calibrator 3C\,286 using a model of the calibrator.
\item{\bf Amplitude \& Phase calibration.}
Gain solutions were determined for the amplitude and phase
calibrators and were used to determine the spectrum of the phase calibrator that was then used to calibrate the target data. Solutions for 3C\,286 used a structural and spectral model and were forced to the spectrum of \cite{PerleyButler2013a}.
\item{\bf Post calibration flagging.}
Another editing pass was done comparing the data with a running mean in frequency.
\item{\bf Rinse and repeat.}
The solutions from the various calibration steps were compared with median values and outliers flagged,  in the solutions and in the accumulated flag table. After the first calibration pass the calibration tables were deleted, the flag table was kept, and the calibration was repeated. 
This ensures that the calibration used well-edited data.
\item{\bf Cross-hand delay calibration.}
The polarized calibrator, 3C\,286, was used to determine
cross-hand delay residuals.
\item{\bf Instrumental polarization calibration.}
The polarization state of the secondary calibrator and the
instrumental polarization were determined using observations over a range of parallactic angle.
\item{\bf Cross-hand phase calibration.}
A cross-hand residual phase spectrum was determined from data on the polarized calibrator 3C\,286 and the polarization model of
\cite{PerleyButler2013b}.
This corrects the polarization angle on the sky.
\end{enumerate}

\subsection{Spectral line data}
For the spectral line data we used a modified version of the VLA scripted pipeline\footnote{https://science.nrao.edu/facilities/vla/data-processing/pipeline/scripted-pipeline} (version 1.3.8) for CASA (version 4.6.0) that has been adapted to work with spectral line data. The calibration pipeline applies preliminary flags, such as flagging data that is not on source, shadowed antennas, and 5\,\% of the channels at both ends of a given sub-band, as these have low signal-to-noise ratios. It determines the absolute flux calibration and derives the delay, bandpass, and gain calibrations. In the default mode, it also automatically flags RFI using the RFI flagging algorithm \texttt{rflag}.  The following modifications were made following the recommendations from the   NRAO: no Hanning smoothing was performed in order to preserve the spectral resolution, the \texttt{rflag} flagging command was only applied to the calibration scans as it can often flag the spectral lines we want to observe, and statistical weights were not calculated as this sometimes affects bright sources, which is a concern for very strong masers. For the pilot region covered in this paper, the complex gain calibrator used was J1804+0101 and the flux calibrators were 3C\,286 and 3C\,48, using the flux density scale of \cite{PerleyButler2013a}. After an initial run of the calibration pipeline, we perform quality checks and manually flag  erroneous data, and rerun the pipeline until satisfied.

\section{VLA data imaging}
\subsection{Continuum data imaging}
The imaging of the continuum data was done with the wide-band, wide-field imaging task \texttt{MFImage} 
in the software package Obit~\citep{Cotton2008}. Instead of a joint deconvolution of all pointings
into a single image, we image each pointing individually and then combine the images into one large mosaic.
For each pointing, we imaged a field of view of 8.4$'$, which corresponds to the full primary beam of the
lower continuum band. Since often there are sources outside the primary beam, which are bright enough to
affect the data, we also use outlier fields at the location of known sources. We place an outlier field
at the location of sources that are within 24$'$ of the pointing center and have a peak intensity of more than
10 mJy/beam. It is possible to  use the NVSS catalog \citep{CondonCottonGreisen1998} to determine the locations of outlier fields. However, since the NVSS
was observed at a lower frequency of 1.4 GHz, there might be sources that are weak in the NVSS, but much
stronger at our frequencies. Hence, we first did a very shallow imaging run of the entire survey using outlier
fields from the NVSS. These images were then used to generate a new catalog of bright sources at 6 GHz that was subsequently used for the outlier fields in the final imaging. 

Using wide-band data (bandwidth synthesis) with adequate spectral resolution provides improved coverage of the u-v plane giving a higher dynamic range as long as the varying source brightness and instrumental response with frequency are adequately accounted for. Wide-band imaging in \texttt{MFImage} is accomplished by dividing the observed total band-pass into frequency bins that are narrow enough not to cause significant problems, which in this case correspond to a fractional bandwidth of 5\%. A frequency variable taper and/or Briggs Robustness factor is needed to produce approximately the same resolution in each frequency bin. Deconvolution using CLEAN proceeds by forming a weighted combined image that is used to select pixels at which CLEAN components are determined. The flux density at each pixel in each frequency bin is recorded and used in the generation of residual images, and  self-calibration and  restoration of CLEAN components. When the CLEAN is complete, each frequency bin is restored
using the components subtracted from it, and a pixel-by-pixel spectrum is fitted correcting for the frequency-dependent primary antenna pattern. Since GLOSTAR observations are part of a mosaic, the primary beam corrections are applied in the linear mosaic produced later.

The three-dimensional aspect of the imaging (the image is flat and the sky is curved) is dealt with using faceting (i.e., tiling the field with smaller images to approximate a curved surface). Each facet is reprojected onto a common grid on a common tangent plane that allows CLEANing all facets in parallel.

The snapshot imaging of GLOSTAR observations produces relatively high sidelobes even with the extended u-v coverage provided by bandwidth synthesis. Therefore, restricting the CLEAN support region is important. The CLEAN window in which components are allowed is automatically adjusted from the statistics of pixel fluxes in each facet in each major cycle, and new portions of the CLEAN window are added as needed.

Radio continuum emission in the Galactic plane has structures on a wide variety of scales; the classical deconvolution into delta-function variant of CLEAN is inadequate. A multi-resolution CLEAN is implemented by generating a set of facet images in each major CLEAN cycle using a set of tapers resulting in a set of resolutions. In each major cycle a decision is made as to which resolution is to be CLEANed. The resultant point or Gaussian components are subtracted from the visibility data and a new set of residual images are generated for the next major cycle. In order to prevent CLEANing too deeply at a given resolution, the depth of each major cycle is restricted\footnote{see Obit Memo 24 for details ftp://ftp.cv.nrao.edu/NRAO-staff/bcotton/Obit/MultRes.pdf}.

Atmospheric phase fluctuations can reduce the dynamic range in pointings with bright emission. A set of phase self-calibrations is triggered if the CLEAN results in a maximum image flux density in excess of a given threshold. A subsequent amplitude and phase self-calibration is triggered by a second threshold.

\begin{figure*}
  \includegraphics[trim=10 100 0 215,clip,width=9cm]{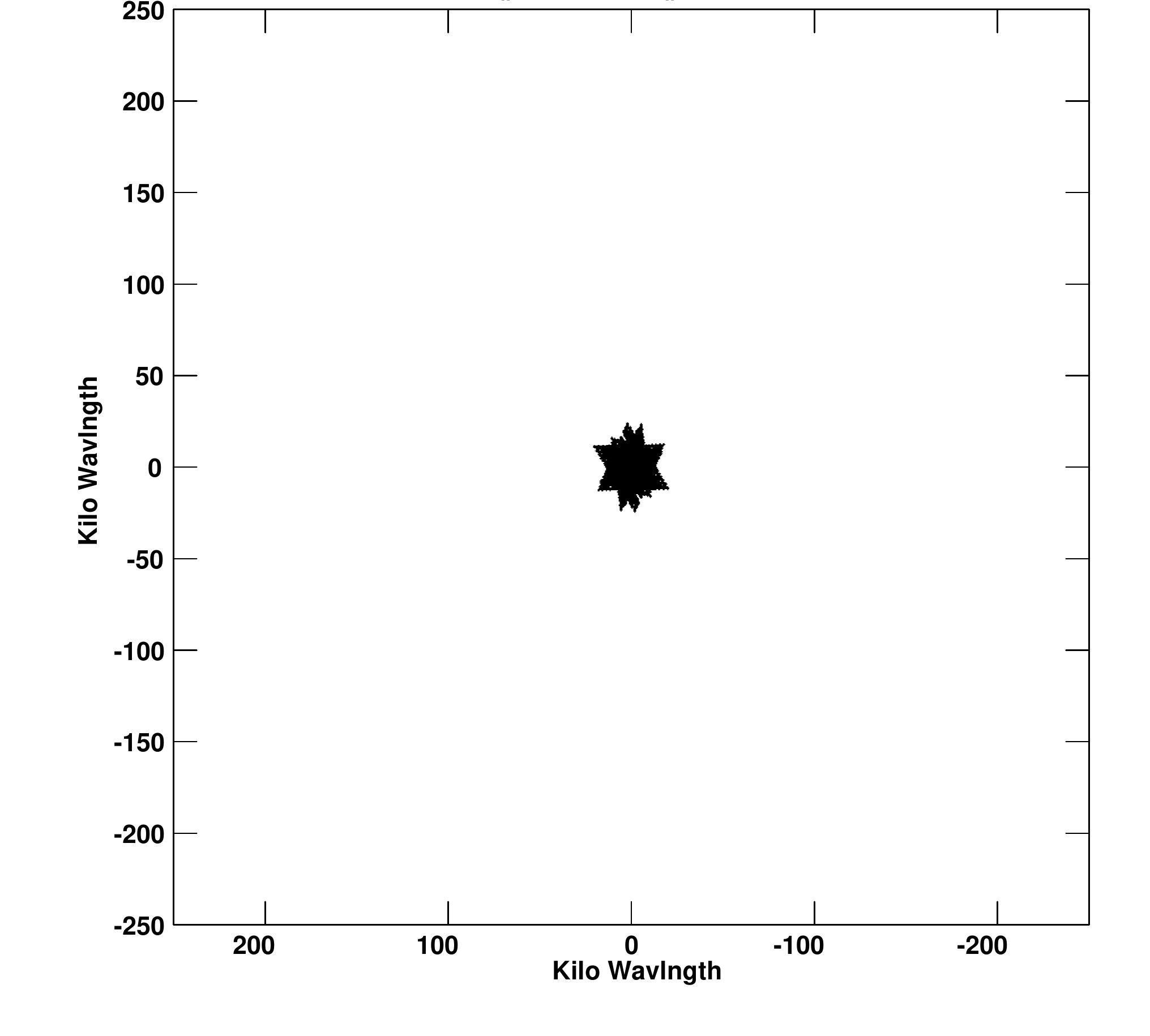}
  \includegraphics[trim=10 100 0 215,clip,width=9cm]{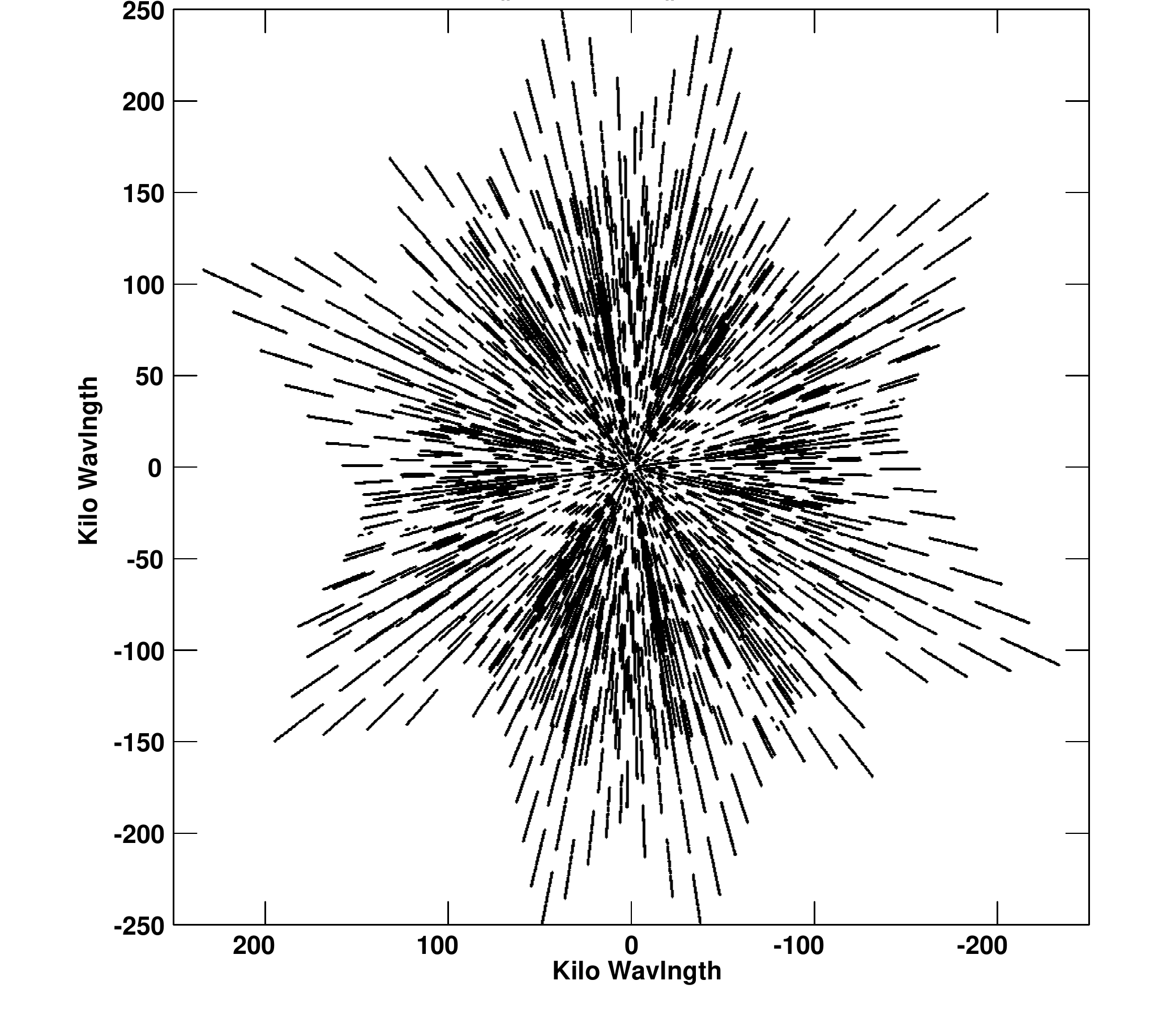}
  \caption{Plots of uv-coverage of a single source at high declination from the D configuration (left) and B configuration (right) observations. While both configurations probe very different ranges, there is significant overlap in the center that allows the combination into a single image. Part of this overlap is due to the large bandwidth covered by the observations.}
  \label{fig:uv-cov}
\end{figure*}

The maximum sensitivity can be obtained by a weighted combination of the overlapping observed pointing images onto a set of images covering the surveyed region. All else being equal, weighting by the square of the antenna
power pattern in a given direction gives the optimum sensitivity. The mosaic formation process for each image plane is given by the summation over overlapping pointing images:
\begin{equation}
  $$ M(x,y)\ =\ {{\sum_{i=1}^{n}A_{i}(x,y)\ I^*_{i}(x,y)}\over{\sum_{i=1}^{n}A^2_{i}(x,y)}}.$$
\end{equation}
Here $A_{i}(x,y)$ is the antenna gain of pointing $i$ in direction $(x,y)$ and $I^*_{i}(x,y)$ is the pointing $i$ observed pixel value
interpolated to direction $(x,y)$ that was multiplied with the antenna gain in the observation and $M$ is the mosaicked image. After combination, the spectrum in each pixel is redetermined. 

\subsubsection{D configuration}

The complex and bright emission in the Galactic plane on various different scales is a challenge, in particular
for snapshot observations. However, the use of two observations separated by a few hours and the large
bandwidth of the observations greatly improve the uv-coverage, even if there are only $\sim$ 15 seconds of
data per pointing. Since the D configuration observations are sensitive to scales up to 4$'$, we used
multi-scale cleaning with three different scales: a delta function and scales of 20 and 40 times the cell size of
the images, respectively. The cell sizes of the individual images were determined from the uv-coverage of the
individual pointings and were typically about 2$''$. Cleaning was done down to a threshold of 0.25 mJy/beam
and a maximum of 2000 iterations.

The individual pointings were all imaged using their intrinsic beam corresponding to the uv-coverage and weighting
used. Hence, the beam sizes are slightly different for each pointing. Before combining all the pointings into one
large mosaic covering the entire pilot region, we convolved all the pointings to a circular beam of $18''$.

\subsubsection{B configuration}

For the B-configuration data we faced different challenges. Since we imaged the same primary beam at about
ten times the resolution, the size of the individual images increased by a factor of 100, resulting in much
slower processing. While most of the large-scale emission in the Galactic plane is resolved out in the
higher resolution images, we found  many artifacts from sources that are just seen on the shortest
baselines of the B configuration (corresponding to $\sim30''$). Since there are only a few baselines that
can see  these extended sources, there is not enough data to image them. Hence, we restricted the uv-range to
baselines of more than 50 k-$\lambda$, which resulted in much cleaner images and fewer artifacts. While there
will always be sources that are visible just on the shortest baselines, the size distributions of the sources
in our survey show that the number of sources drops significantly above a size of 30$''$ \citep{Medina2019}.
Since the overwhelming majority of sources in the resulting images are now very compact, there is no need for
multi-scale cleaning. The B configuration images are much less complex than the lower resolution maps.
Therefore, it was possible to reduce the number of clean iterations to 250 per pointing. The threshold for
cleaning was also 0.25 mJy/beam, and the cell sizes in the images were $0.2''$.

The individual pointings were again imaged using their intrinsic beam corresponding to the uv-coverage and weighting
used. Hence, the beam sizes are slightly different for each pointing. Since the images of the B-configuration data
have about 100 times more pixels than the D-configuration images, it was not practical to combine all the pointings
of the pilot region into a single mosaic. As a compromise, we combined pointings into eight $2^\circ \times 1^\circ$ strips
across the Galactic plane (roughly corresponding to the individual observations, but also including pointings from the
edges from the neighboring observations). Before combining the pointings they were convolved to a  circular beam of $1.5''$.

\subsubsection{Combination of D and B configurations}

There is a significant overlap in the baselines between the D configuration (35 m -- 1030 m) and the
B configuration (210 m -- 11100 m). Hence, it is possible to combine the two data sets and image them together. The uv-coverage for a single high-declination source from the D- and B-configuration data are shown in Fig~\ref{fig:uv-cov}. The resulting images have a reasonably good sensitivity to extended structures, but still  a much higher resolution than only the D configuration images. While adding data from the intermediate C configuration with baselines between $35m$ and $3400m$ would be better, the additional observing time is difficult to justify.

The combination of D- and B-configuration data is the most time consuming since it still requires a large number of pixels and multi-scale cleaning. Since the range of different scales is even larger than in the D configuration images, we now allow for four different scales: a delta function, and scales of 20, 50, and 100 times the pixel size of $0.5''$. A test with  more scales made the imaging even slower, but did not change the final image significantly. The maximum number of iterations and the clean threshold were set to 2000 and 0.25 mJy/beam, respectively.  Before combining the pointings they were convolved to a circular beam.  Since these combined images are mainly useful for structures with intermediate sizes, we used a beam of $4''$ for the convolution.

\subsubsection{Astrometric accuracy}
\label{sect:astrometric_accuracy}

As already discussed in \cite{Medina2019}, the astrometric accuracy of the D-configuration observations is about $2''$. For the B-configuration observations, \cite{Ortiz2021} estimated position errors of $<0.2''$ by comparing methanol maser positions in the Cygnus~X region with known positions from VLBI observations from \cite{Rygl2012}. A more detailed discussion about the astrometric accuracy of the B-configuration data will be presented in forthcoming publications with the catalogs of these data.

\subsection{Spectral line data imaging}

\subsubsection{Dirty image}
\label{sect:dirtyImage}

Once  the calibration, RFI removal, and continuum subtraction is complete, we move onto the imaging and source finding stage. This is done using the CASA software;  one of the main reasons it is preferred over other software is due to its mosaicking task that can do joint deconvolutions of different fields.

We first produce what we call a ``dirty'' image of the D-configuration data, which is essentially the brightness distribution convolved with what is called the synthesized--dirty beam or point spread function (PSF). This image uses only the sampled visibilities and cannot recreate the entire sky brightness distribution without reconstruction techniques. These techniques are implemented via the CLEAN algorithm.
Ideally, we would run the CLEAN algorithm over the entire data cube; however, for a
$1^{\circ}\times1^{\circ}$ cube spanning at least 1000 channels, this can take around 3 weeks
to complete. We opt instead to produce these dirty images, which take around 2 days for a 
cube of the same size.

In CASA the \texttt{tclean} task is implemented for the imaging of the dirty image with the following parameters: \texttt{niter}=0, since we do not  deconvolve the image as we are not performing a full CLEAN at the moment; cube size (\texttt{imsize}) of $2500\times2500$ pixels, and  pixel size (\texttt{cell}) of 2.5\arcsec. The number of channels (\texttt{channels}) and the rest frequency (\texttt{restfreq}) are set dependent on the transition. The \texttt{weighting} parameter is set to ``natural'' to  better   capture weaker sources;  natural weighting has the best sensitivity, while ``uniform'' weighting gives better resolution.

The 1$\sigma$ rms noise in the line free channels of the dirty cubes is found to be $\sim$20~mJy~beam$^{-1}$, which is consistent with the noise estimated using the VLA Exposure Calculator for both the methanol and formaldehyde lines. We note that the expected noise level 
for a total of 15 second integration time in the D configuration with a channel width 
of $\sim$3.9 kHz is actually $\sim$40~mJy~beam$^{-1}$, but
since each field is covered by six neighboring fields the final sensitivities will be
at least a factor of two better ($\sim$20~mJy~beam$^{-1}$), as mentioned before.

We find that weak and isolated masers can be detected already from the dirty image, meaning
that it is not necessary to CLEAN the entire image. Only weak masers that are in the vicinity
of strong masers in both the spatial and velocity domains can possibly be missed. 
It is thus much more efficient  to search for these dirty cubes first and compile a 
preliminary catalog of detections. The next step is to cut out smaller regions encompassing
the detections and CLEAN them to search for weaker detections that were in the vicinity
of stronger detections. A detailed description of the source finding algorithm for the masers that was used to find the sources will be included in an upcoming paper (Nguyen et al., in prep.).

\subsubsection{Final CLEAN}
\label{sect:CLEAN}
Patches of $0.23^\circ \times 0.23^\circ$ using on average 16 pointings centered on the sources detected in the dirty images were split out to make smaller visibility data sets for CLEANing. The 
cell size and/or pixel size was retained at 2.5$\arcsec$ for the D configuration, and the imaging was restricted to the velocity ranges over which a significant signal was detected. This produced images with spatial extents of $350\times350$ pixels and spectral extents of $\sim$18-25~\kms\, centered on the peak velocity. A CLEANing threshold of 60~m\jyb\ was chosen ($\sim$3$\sigma$) and the parameter
niter (number of minor CLEAN cycles) was set to  1000, 5000, or 10000 depending on the strength of the maser involved, with \texttt{niter} = 1000 to CLEAN patches
containing weak masers to \texttt{niter} = 10000 for patches containing very strong sources. These iteration values were found to optimize the automatic cleaning for many sources to maximize image fidelity while avoiding overcleaning.
The weighting parameter was set to \texttt{uniform} since it gives a better resolution, thus giving better positional accuracy of the source.

Additionally,  B configuration images were also produced for the methanol maser data. With a higher angular resolution of $\sim1\arcsec$, to image and CLEAN an area of  similar size to the D-configuration data would take significantly more time. However, thanks to the benefit of first studying  the D-configuration data, we made smaller images ($0.1^\circ \times 0.1^\circ$) using the B-configuration data centered on detected features from the D-configuration data, thereby efficiently using our available computing resources. We used a pixel size of 0.3$\arcsec$, resulting in images with spatial extents of $1200\times1200$ pixels and similar spectral extents to those above. Similar CLEANing iteration parameters were used depending on the individual source intensity.

\subsubsection{Radio recombination line imaging}
To increase the signal-to-noise ratio of RRL emission, we observed multiple RRLs with the goal to stack them in velocity. We observed seven RRLs (Table~\ref{correl}), which were chosen to avoid radio frequency interference (RFI) and due to bandwidth limitations. We inspected the calibrated data for RFI with   automatic tools, such as {\tt rflag} in CASA, and manually, and flagged data that were corrupted. Since we found that the H96$\alpha$ line is strongly affected by RFI, we decided to exclude this transition in the following.

The lines were imaged individually in fields with dimensions $\Delta l = 1$\degree and $\Delta b = 2$\degree. We included in the imaging pointings from adjacent scheduling blocks to provide uniform sensitivity across the edges of the images, and to include sources whose sidelobes may significantly affect the image quality and need to be accounted for during the imaging process. We subtracted the continuum in the uv-domain from line-free channels, which are selected from uniform line-free velocity ranges. The data were imaged and deconvolved with a threshold of several times the noise level using \texttt{tclean} in CASA (version 5.7.0), with the sub-parameters {\tt imagermode=``mosaic''} and {\tt gridder=``mosaic''}. By setting {\tt restoringbeam=``common''} the residuals of all channels in a spectral cube were subsequently smoothed and the image restored with a common beam. The original bands span 8\,MHz with 128 channels, yielding a velocity resolution of 2-4\,\kms\ (see Table~\ref{correl}). The images were gridded in velocity bins of 5\,\kms\ with a velocity range from -150\,\kms\ to 150\,\kms.

In this paper we show the RRL emission for two regions, the massive star-forming region G29.92-0.03, also known as W43-south, and  the \ion{H}{ii} region G28.60-0.37. To optimize imaging and angular resolution, the data were re-imaged for these objects in two cubes with dimensions of $\Delta l = 0.2$\degree and $\Delta b = 0.2$\degree, and all lines smoothed to an angular resolution of 25\arcsec\ before stacking. The images were rotated into Galactic coordinates. The noise levels of the smoothed images are in the  range   5--10\,m\jyb\, with the noise of the stacked images of $\sim3$\,m\jyb\ at an average frequency of 5.3\,GHz. A full description of the RRL data will be given in a forthcoming publication.

\section{Effelsberg observations and imaging}
\subsection{Observations}
The GLOSTAR single-dish observations of the pilot region were performed with the 100-m telescope at Effelsberg, Germany,\footnote{The 100-m telescope at Effelsberg is operated by the Max-Planck-Institut f{\"u}r Radioastronomie (MPIFR) on behalf of the Max-Planck Gesellschaft (MPG).} from 2019 January to September. We used the S45mm receiver installed at the secondary focus as front end. The S45mm receiver is a 4.5 cm broadband receiver with two orthogonal linear polarizations. We simultaneously employed two different kinds of back ends:  SPEctro-POLarimeter (SPECPOL) and fast Fourier transform spectrometers (FFTSs: \citealp{Klein2012}). SPECPOL is used to obtain the Stokes parameters (I, Q, U, V). SPECPOL delivers two bands covering 4--6 GHz and 6--8 GHz. Each band provides 1024 channels, resulting in a channel width of 1.95~MHz. The phase shifts between different channels in Stokes parameters were fixed by the ADC synchronization procedure before our survey observing sessions. FFTSs deliver two low spectral resolution bands and two high spectral resolution bands that are used for spectral studies. The two low spectral resolution bands cover 4--6.5 GHz and 5.5--8 GHz. Each band provides 65536 channels, resulting in a channel width of 38.1 kHz, corresponding to a velocity spacing of 2.2 km~s$^{-1}$ at 5.25 GHz. The two high spectral resolution bands are designed to study H$_{2}$CO (1$_{1,0}$--1$_{1,1}$) and CH$_{3}$OH (5$_{1}$--6$_{0}$ A$^{+}$). These two bands have bandwidths of 200~MHz. Each band provides 65536 channels, resulting in a channel width of 3.1 kHz, corresponding to a velocity spacing of 0.14 km~s$^{-1}$ at 6.7 GHz (for a selection of the observed spectral lines, see Table~\ref{tbl:effobs}).

\begin{table}[h]
\centering
\caption{Selection of spectral lines covered with the GLOSTAR Effelsberg observations}
\label{tbl:effobs}
\begin{tabular}{lrrrrrrc}
Line & Freq. &  $\Delta v$ (nat.)$^{a}$ & $\Delta v$$^{b}$& $\theta$$^{b}$ & RFI$^{c}$\\
     & [MHz]     &    [\kms] &    [\kms]  & [\arcsec] &  \\
\hline
\hline
 H94$\alpha$ & 7792.871 & 1.47 &5.00 &94.6 &$\cdot$ \\ 
 H95$\alpha$ & 7550.614 & 1.52 &5.00 &97.6 &$\cdot\cdot$ \\ 
 H96$\alpha$ & 7318.296 & 1.56 &5.00 &100.7 &$\cdot\cdot$ \\ 
 H97$\alpha$ & 7095.411 & 1.61 &5.00 &103.8 & \\ 
 H98$\alpha$ & 6881.486 & 1.66  &5.00 &107.1 &$\cdot$ \\ 
 H99$\alpha$ & 6676.076 & 1.71  &5.00 &110.4 & \\ 
 H100$\alpha$ & 6478.760 & 1.77 &5.00 &113.7 & \\ 
 H101$\alpha$ & 6289.144 & 1.82 &5.00 &117.2 & \\ 
 H102$\alpha$ & 6106.855 & 1.87 &5.00 &120.7 & \\ 
 H103$\alpha$ & 5931.544 & 1.93 &5.00 &122.6 & \\ 
 H104$\alpha$ & 5762.880 & 1.99 &5.00 &126.2 &$\cdot$ \\ 
 H105$\alpha$ & 5600.550 & 2.04 &5.00 &129.8 &$\cdot$ \\ 
 H106$\alpha$ & 5444.260 & 2.10 &5.00 &133.6 & \\ 
 H107$\alpha$ & 5293.732 & 2.16 &5.00 &137.3 &$\cdot\cdot$ \\ 
 H108$\alpha$ & 5148.703 & 2.22 &5.00 &141.2 &$\cdot$ \\ 
 H109$\alpha$ & 5008.923 & 2.28 &5.00 &145.2 &$\cdot$ \\ 
 H110$\alpha$ & 4874.157 & 2.35 &5.00 &149.2 &$\cdot$ \\ 
 H111$\alpha$ & 4744.183 & 2.41 &5.00 &153.3 &$\cdot$ \\ 
 H112$\alpha$ & 4618.789 & 2.48 &5.00 &157.4 &$\cdot$ \\ 
 H113$\alpha$ & 4497.776 & 2.54 &5.00 &161.7 &$\cdot$ \\ 
 H114$\alpha$ & 4380.954 & 2.61 &5.00 &166.0 &$\cdot$ \\ 
 H115$\alpha$ & 4268.142 & 2.68 &5.00 &170.4 &$\cdot\cdot$ \\ 
 H116$\alpha$ & 4159.171 & 2.75 &5.00 &174.8 &$\cdot\cdot$ \\ 
 H117$\alpha$ & 4053.878 & 2.82 &5.00 &179.4 &$\cdot\cdot$ \\ 
 H$_2$CO & 4829.660 & 0.19 &0.20 &150.6 &$\cdot$ \\ 
 H$_2^{13}$CO & 4593.088 & 2.49 &2.50 &158.3 & \\ 
 CH$_3$OH & 6668.519 & 0.14 &0.15 &110.5 & \\ 
 OH$^{d}$ & 4660.241 & 2.46  &2.46 &156.0 &$\cdot\cdot$ \\ 
 OH$^{d}$ & 4750.660 & 0.19  &0.20 &153.0 &$\cdot$ \\ 
 OH$^{d}$ & 4765.562 & 0.19  &0.20 &152.6 &$\cdot\cdot$ \\ 
 OH$^{e}$ & 6030.748 & 1.90  &2.00 &122.2 & \\ 
 OH$^{e}$ & 6035.093 & 1.90  &2.00 &122.1 & \\ 
 OH$^{e}$ & 6016.748 & 1.90  &2.00 &122.5 & \\ 
 OH$^{e}$ & 6049.084 & 1.89  &2.00 &121.8 & \\ 
 
\hline
\hline
\end{tabular}
\tablefoot{$(a)$ Native velocity channel width. The actual spectral resolution corresponds to twice the channel width as the data is smoothed in the process of the LSR velocity correction. $(b)$ Channel width and FWHM of the gridded data cubes. $(c)$ RFI/Artifact contamination of the pilot region data without any removal: ``$\cdot$'' indicates medium contamination, possibly removable by excision techniques, while ``$\cdot\cdot$'' indicates ubiquitous, strong contamination. $(d)$ Transitions in OH ${}^2\Pi_{1/2}(J=1/2)$: F=$0^--1^+$, F=$1^--1^+$, F=$1^--0^+$. $(e)$ Transitions in OH ${}^2\Pi_{3/2}(J=5/2)$: F=$2^--2^+$, F=$3^--3^+$, F=$2^--3^+$, F=$3^--2^+$.}
\end{table}

The pilot region is divided into cells of 0.2$^\circ$ ($l$) $\times$ 2$^\circ$ ($b$) and 2$^\circ$ ($l$) $\times$ 0.2$^\circ$ (b). Each cell was observed in the on-the-fly (OTF) mode with a scanning speed of 90$''$ per second with a step of 30$''$ between OTF subscans in order to conform to the Nyquist sampling theorem. Each dump consists of two phases, for one of which the calibration signal is switched on. The sampling time is 125 ms for each phase. The pilot region was observed twice, along the Galactic longitude and latitude, in order to reduce the scanning effects and allow basket weaving (e.g., \citealp{Winkel2012}). 

Pointing and focus was checked on strong continuum sources at 6.7 GHz. NGC\,7027 was used as flux calibrator, while 3C\,286 was used as polarization calibrator. The half power beam width (HPBW) was 140$''$, at 5 GHz. The typical system temperature was 28--42 K. Almost all observations were carried out during the night to avoid the influence of the strong solar emission from the telescope sidelobes. 

\subsection{Calibration}

\begin{figure*}
\includegraphics[width=19cm]{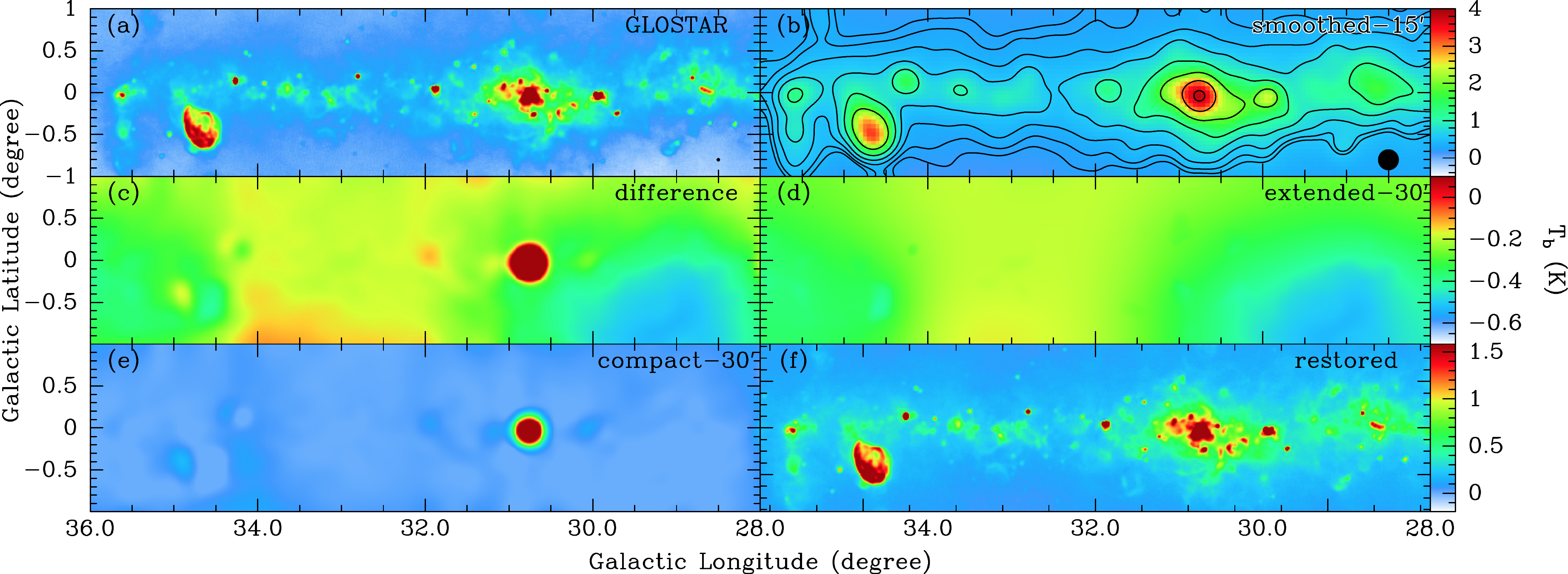}
\caption{Zero-level shift correction for the Effelsberg data. (a) Effelsberg 4.8 GHz continuum image. (b) Urumqi 6 cm continuum emission overlaid with the Effelsberg 4.8 GHz continuum contours. Both data sets have been convolved to an angular resolution of 15\arcmin. The contours start at 50 mK, and each contour is twice the previous one. (c) Difference map derived by subtracting the Urumqi 6 cm continuum emission from the Effelsberg 4.8 GHz continuum map in Fig.~\ref{fig:bgf}b. (d) Extended emission component of Fig.~\ref{fig:bgf}c derived by the background filter method. (e) Compact emission component derived by the background filter method (f) Restored Effelsberg 4.8 GHz continuum image.}
\label{fig:bgf}
\end{figure*}

The polarized calibrator 3C\,286 was used to calibrate the flux density and polarization of our radio continuum data. Based on the Effelsberg long-term monitoring
observations (A. Kraus, priv. comm.), the flux density of 3C\,286 follows the relationship

\begin{equation}
\begin{split}
 {\rm log}_{10}(S_{\nu} [{\rm Jy}]) = 1.11578+ 0.494671\times\log_{10}\nu [{\rm MHz}]\\
-0.151995\times ({\rm log}_{10}\nu [{\rm MHz}])^{2} \;,
\end{split}
\end{equation}
and has a nearly constant linear polarized angle of 33\degree\,and a linear polarization degree of 11\% at the observed frequencies (4--8 GHz). Based on these values, we obtained the M{\"u}ller matrix by observing 3C\,286 at different parallactic angles. The derived M{\"u}ller matrix is then applied to our data in order to correct the instrumental effects of the receiver. In addition, the parallactic dependence of the polarization measurements are corrected.

Data reduction was carried out with the NOD3 software package \citep{MuellerKrauseBeck2017}. Strong RFI is present across the observed frequencies. Thanks to the newly installed SPECPOL, we were able to select specific frequencies while filtering out RFI-affected channels. This resulted in two continuum images  centered at 4.8 GHz and 7.0 GHz, respectively. The remaining RFI was  flagged by data editing in NOD3. For each subscan, a linear baseline is subtracted in order to correct gain drifts. For each field, the coverage scanned in the $l$ and $b$ directions were combined with the basket weaving technique, which significantly reduces the scanning effects and artifacts.

Since we only observed the Galactic plane within $|b|\leq1.0$\degree, all maps have a relative zero-level shift due to the baseline subtraction. These zero-level shifts result in negative values in the reduced map (see Fig.~\ref{fig:bgf}a). Here we used the Sino-German 6 cm data from the Urumqi telescope \citep{SunHanReich2007,SunReichHan2011} to restore the extended emission at 4.8 GHz because the effective center frequency is nearly identical. We first applied high cuts to both images to avoid the contamination from the convolution of strong compact sources that may spread emission to extended emission in the convolved images. The high cut value was set to be 15 K and 4 K for the Effelsberg and Urumqi images, respectively. We then convolved both images to an angular resolution of 15\arcmin (see Fig.~\ref{fig:bgf}b), and obtained the difference of the two images (see Fig.~\ref{fig:bgf}c). In order to only restore the large-scale emission, we applied the background filtering method to the difference map \citep{1979A&AS...38..251S}. This method is used to avoid the contributions of the Urumqi data on small scales that can be affected by pointing errors and slightly deformed shapes of compact sources. We used a filtering beam size of 30\arcmin\,and 50 iterations to decompose the difference map until the compact and extended emission are well separated (see Figs.~\ref{fig:bgf}d--\ref{fig:bgf}e). Then the extended component of the difference map (Fig.~\ref{fig:bgf}d) was used to correct the zero-level shift in the original Effelsberg 4.8 GHz continuum data, which leads to the restored data set (see Fig.~\ref{fig:bgf}f). Since there is no available public data that can be used to correct the zero-level shift at 7 GHz, we derived the extended emission at 7 GHz using the data from the Effelsberg 11 cm survey \citep{ReichFuerst:1990aa} and the Urumqi 6 cm survey \citep{SunHanReich2007,SunReichHan2011}. These data were convolved to an angular resolution of 15\arcmin in order to derive the spectral index of the extended emission. The 7 GHz extended emission was then calculated by interpolation with the derived spectral index. Similarly, we calculated the difference between the derived and observed 7 GHz data at a common angular resolution of 15\arcmin, and we used the same method to restore the extended emission at 7 GHz. We also note that the extended emission at 7 GHz is derived from the interpolation of images at only two frequencies. Thus, the uncertainties in the extended emission are greater in the 7 GHz continuum data than in the 4.8 GHz continuum data.

\begin{figure}
\includegraphics[width=9cm]{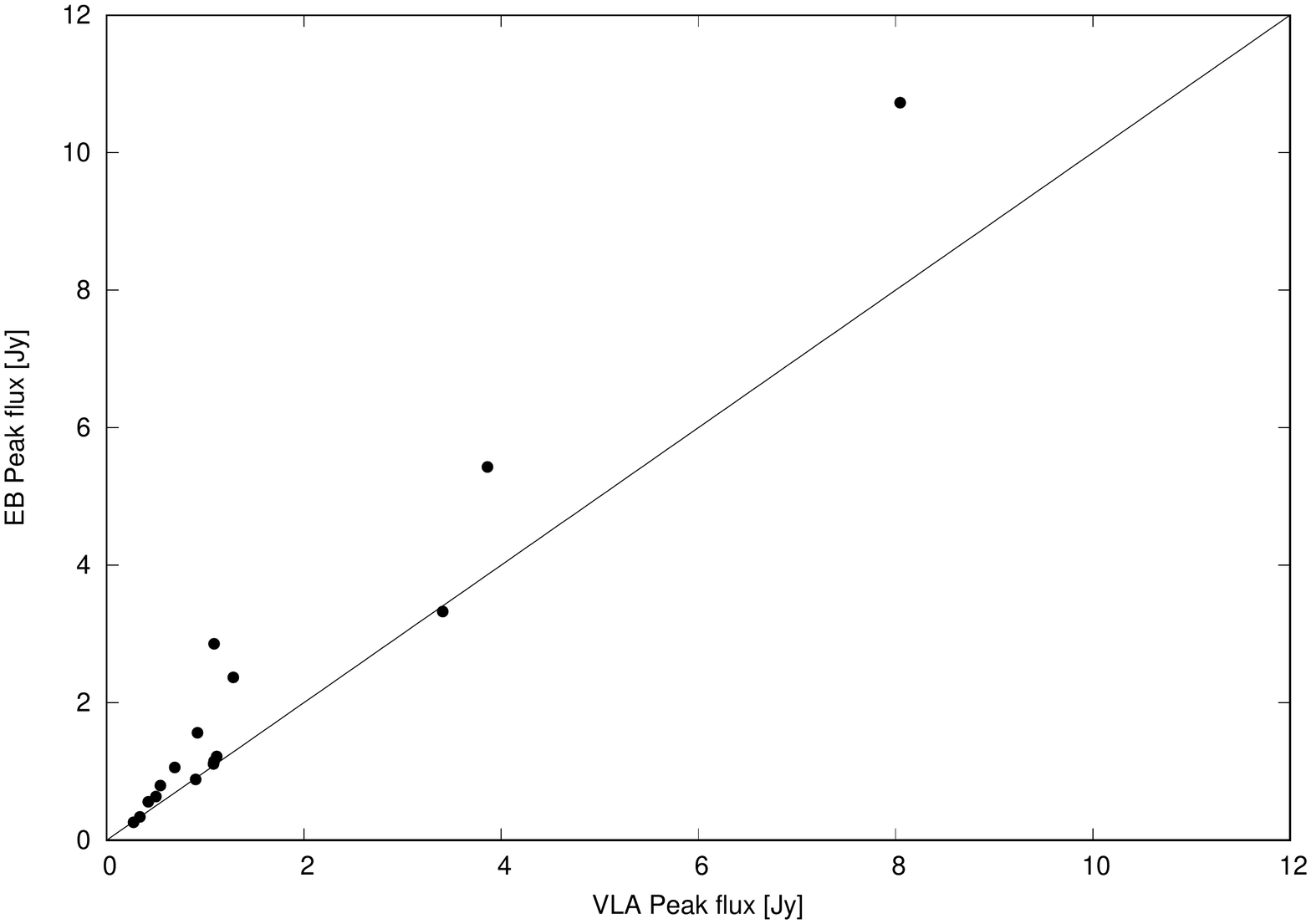}
\caption{Comparison of peak fluxes of several sources for the Effelsberg and VLA images. The solid line denotes a 1:1 ratio.}
\label{fig:fluxes}
\end{figure}

The calibration of the spectral line data follows the recipes described in \cite{Kraus:2009aa}, with several improvements to account for the frequency dependence of receiver gain, system and noise diode temperatures and atmospheric effects (see \citealp{Winkel2012b}). In particular, atmospheric opacity is corrected for using a newly developed water vapor radiometer. For absolute flux calibration, we used spectroscopy pointings on the unpolarized calibrator NGC 7027 in comparison with a model. The two wide bands of 2.5 GHz were split into several sub-bands for calibration. The two 200 MHz narrow bands were calibrated independently. The spectra were converted to main beam temperature using a main beam efficiency of $\eta_{\rm mb}$$\sim$$0.65$.

\begin{figure*}
\includegraphics[width=19cm]{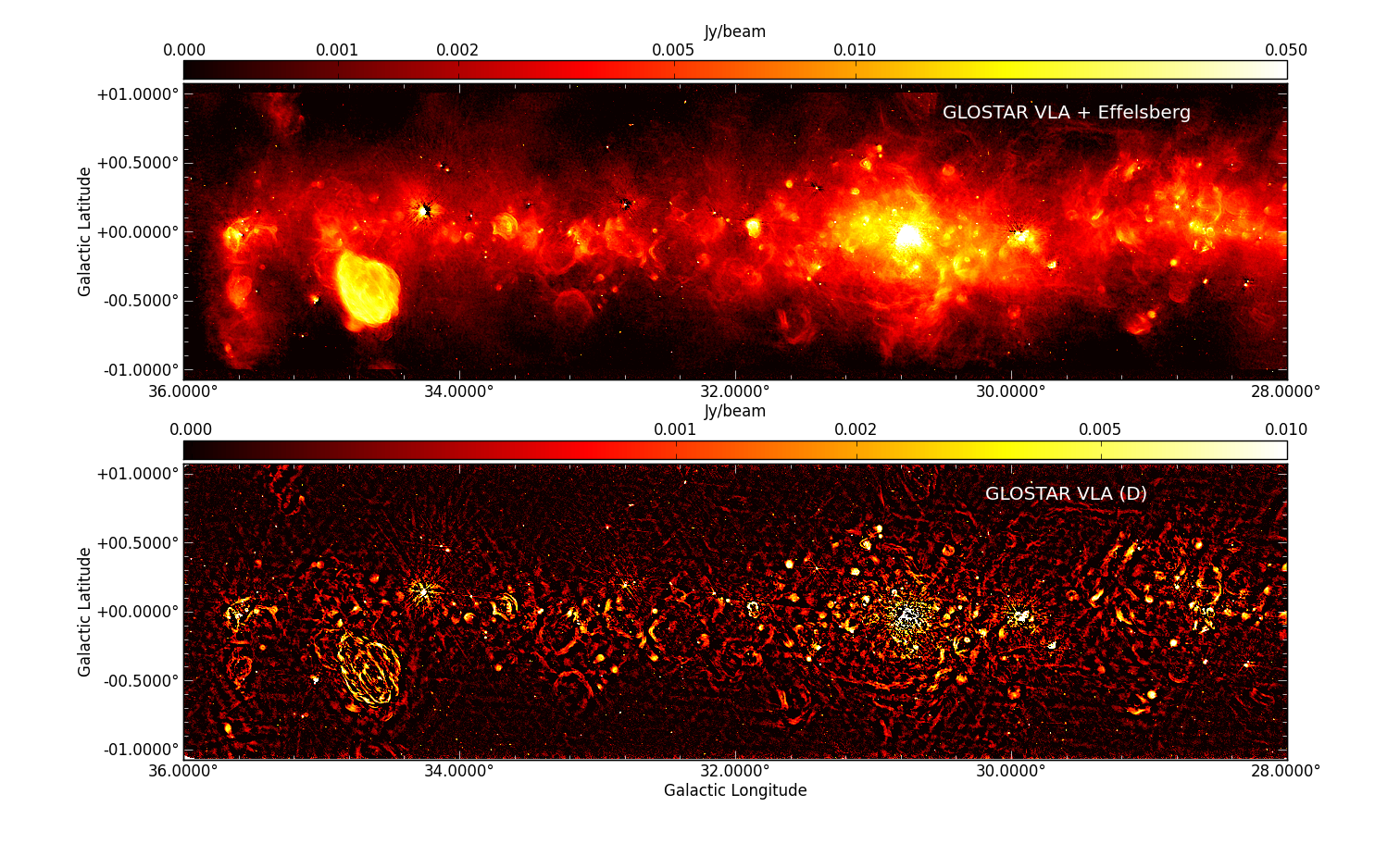}
\caption{Radio contiuum image of the pilot region in the range $28^\circ<l<36^\circ$. {\bf Top:} The combination of the VLA D configuration and the Effelsberg single-dish images. {\bf Bottom:} D configuration VLA image of the full continuum of the same longitude range already shown in \cite{Medina2019}.}
\label{fig:cont}
\end{figure*}

The frequency range  4--8\,GHz contains a variety of spectral lines, all of which are simultaneously covered with the WFFTS back ends. We focused on a subset of these, which are very interesting for the study of the ISM and are likely to be detected. We note that not all of them are usable due to the contamination by artifacts such as RFI. An overview of the lines is given in Table~\ref{tbl:effobs}, together with the native channel width and typical RFI contamination before any RFI removal. After calibration we subtracted a first-order baseline for each spectral line using line-free channels that are offset in velocity from the main Milky Way emission. The final mapping of each line was done in GILDAS\footnote{\url{https://www.iram.fr/IRAMFR/GILDAS/}} with CLASS \citep{Pety:2005aa}. We used the standard gridding kernel of 1/3 of the beam size, which yields slightly lower angular resolution in the final images than the raw telescope beam. We grid the data from both longitude and latitude scans together. The noise levels vary across the maps due to different elevations and weather conditions. For molecular transitions, we resampled the final maps with channel widths close to the original resolution, while the actual velocity resolution was  a factor of two coarser since the data was smoothed with a Gaussian kernel in the process of regridding to the LSR rest frame. For the RRLs, due to their broader intrinsic line widths and to obtain independent channels at all frequencies, we resampled all lines at 5\,\kms. Details of the data cubes of all gridded lines are given in Table~\ref{tbl:effobs}. Given the wide bandwidth of the FFTS back ends, velocity coverage is continuous. We typically grid a wide enough velocity range to cover Galactic emission and to verify that the region for baseline subtraction is not affected by RFI.
Stacking multiple RRLs in velocity further increases the signal-to-noise ratio. Using approximately seven RRLs without further flagging of RFI, smoothing all lines to a common angular resolution of 180\,\arcsec, and using the mean to combine the maps at each position and velocity channel yields an RMS of typically 0.008\,K at 180\,\arcsec\ resolution in a channel  5\,\kms\ in width.

\subsection{Combination of VLA and Effelsberg data}

Combining the Effelsberg data and the VLA D-configuration continuum data is not straightforward, due to the different frequency coverages. While the VLA observations cover two 1 GHz bands, the Effelsberg observations include the full range from 4 to 8 GHz. Furthermore, the RFI situation at the two locations is very different. Hence, it is not possible to compute images averaged over frequency and then combine the two averaged images. Instead, we used the two images obtained from the Effelsberg data with frequency ranges of 4-6 GHz and 6-8 GHz and average frequencies of 4.8 GHz and 7 GHz, respectively. With the multi-frequency data from the VLA data, we generated two images corresponding to the same frequencies. Then we used the task {\it Feather} in Obit \citep{Cotton2017} to generate two combined images of VLA and Effelsberg at   low and high frequencies. Finally, these two images were combined pixel-by-pixel to create the final image and a spectral index map. The combination of the spectral line data from the VLA and Effelsberg will be done in the future.

\begin{figure*}
\includegraphics[width=19cm]{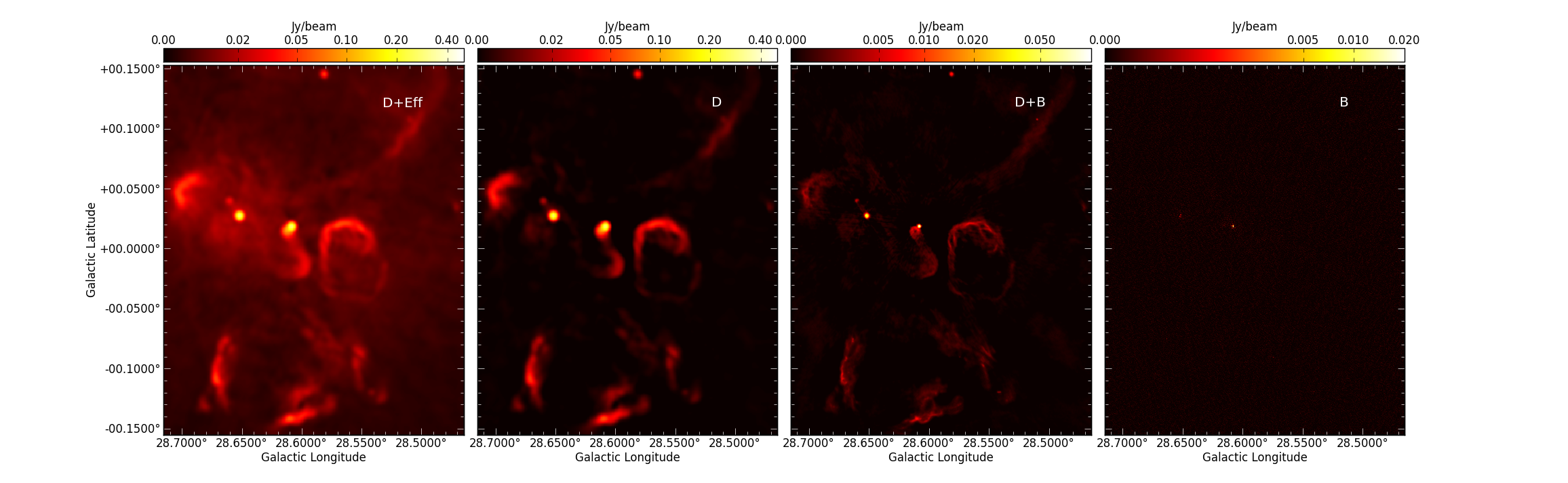}
\caption{Comparison of a smaller region between the D configuration and Effelsberg combination (left), the D configuration only  (middle left), the combination of D and B configurations (middle right), and the B configuration only  (right).}
\label{fig:all_conf}
\end{figure*}

Comparing the flux scales between the Effelsberg and VLA observations is difficult due to the different spatial resolutions.  The Effelsberg images obviously pick up more extended emission, which is filtered out by the VLA. However, the Effelsberg data and the Urumqi 6 cm data agree to within 2\%, and the flux calibration of VLA observations is typically also good to at least 5\%. \cite{Medina2019} found  good agreement between the  GLOSTAR and CORNISH/MAGPIS fluxes. To compare the scales, we smoothed the VLA image at 4.8 GHz to the beam size of the Effelsberg observations and then fitted Gaussians to several sources in the entire region that are relatively isolated (although there is always some diffuse emission in the Effelsberg images). The results are shown in Fig.~\ref{fig:fluxes}. While some sources have higher Effelsberg fluxes, as expected if there is extended emission, there is no clear systematic offset between the two data sets. Hence, no flux scaling was performed before combining the images.

\section{Results}
The continuum and spectral line data at various resolutions allows us to address a large variety of different science cases. It would be well beyond the scope of this paper to discuss these in detail. Hence, this section gives only a small overview of some of the science that can be done with this data. This will be investigated in much more detail in forthcoming papers. We plan to release all continuum images (D+Effelsberg, D, D+B, and B) as well as continuum and spectral line source catalogs and spectra of the detected lines in the future on our survey website.~\footnote{https://glostar.mpifr-bonn.mpg.de}

\begin{figure*}
\includegraphics[width=18cm]{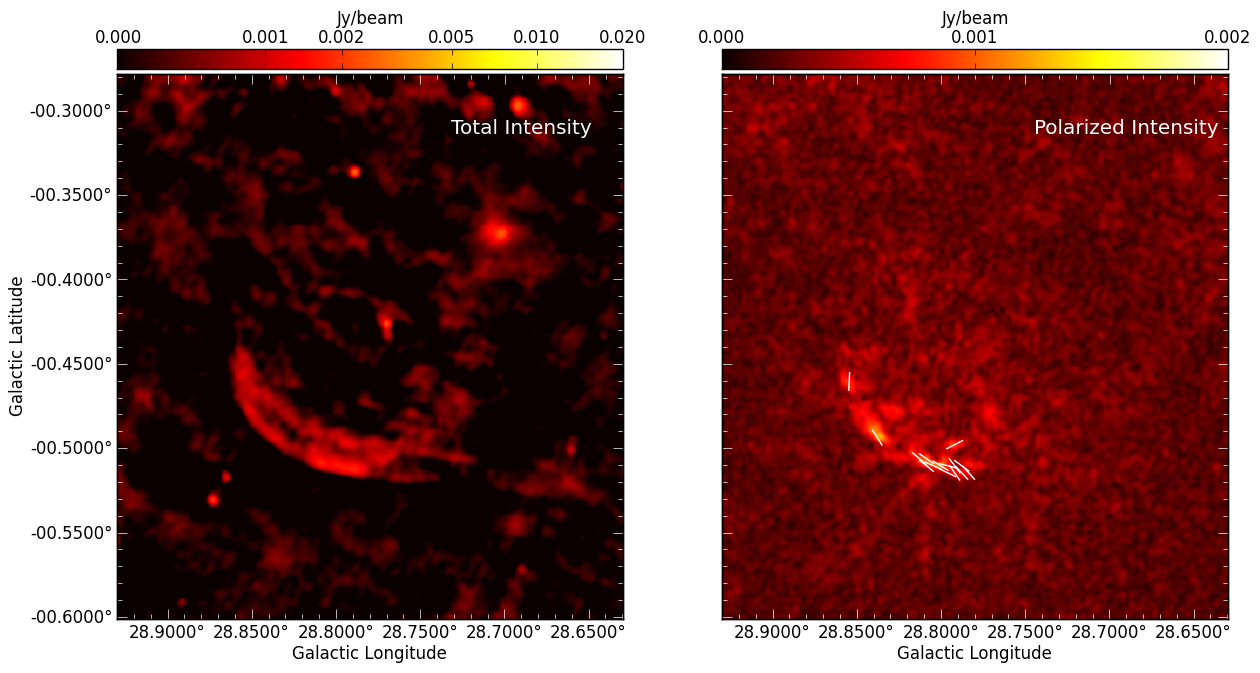}
\caption{Continuum images of the supernova remnant G28.78-0.44 from the VLA D-configuration data in total intensity (left) and polarized intensity with electric field vectors (right).}
\label{fig:G28.78}
\end{figure*}

\subsection{Continuum}
The full continuum image of the pilot region covering the range from 28$^\circ<l<36^\circ$ is shown in Fig.~\ref{fig:cont}. Here we show the combined image using the VLA D configuration and the Effelsberg data (top) as well as the  VLA-only image (bottom). As expected, the addition of the Effelsberg single-dish data helps to resolve many of the missing flux problems seen in the D configuration-only images. However, a few artifacts remain, mainly near very bright and compact sources.

In the D configuration image we find a total of 1575 discrete radio sources and 27 large-scale structures. From these sources, 231 continuum sources were classified as H{\sc ii} regions, 37 as ionization fronts, and 46 as planetary nebulae. The catalog of sources is published in \cite{Medina2019}. In the B configuration images, we detect about 1400 sources above 7$\sigma$ in total (Dzib et al. in prep). The similar detection rate to the D configuration images can be explained by  lower noise in regions with extended emission in the D configuration images, by sources that split into multiple sources at a higher spatial resolution, while extended sources are completely resolved out. In the same region the CORNISH survey with the same spatial resolution as our B configuration images finds a total of 353 sources \citep{PurcellHoareCotton2013}.

In addition to many sources that are located in the Milky Way, there is also a large number of extragalactic sources in the survey. While these sources were not the main focus of the survey, they are an interesting by-product (e.g., \citealt{Chakraborty2020}).  

\begin{figure*}
\begin{center}
    \includegraphics[width=\linewidth]{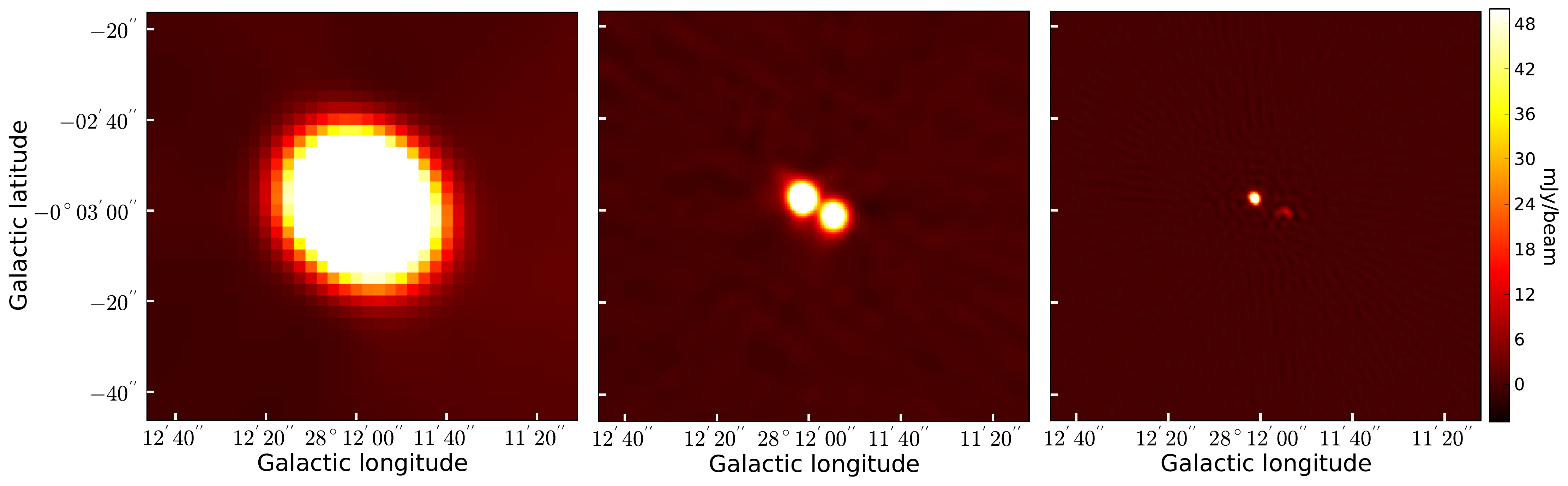} 
\caption{From left to right images from D, D+B, and B configurations  of G28.2-0.04.}
\label{fig:HCHIIG28}
\end{center}
\end{figure*}

In Fig.~\ref{fig:all_conf} we show the details of a smaller region on all available scales, starting with the Effelsberg and D configuration combination down to the image using only the B configuration that contains only the most compact sources. While some of the extended emission is resolved out in the combined images, the higher resolution shows finer details of the sources.

\subsubsection{Supernova remnant G28.78-0.44}

Though the non-thermal emission is faint in C band compared to lower frequencies, the high sensitivity of the GLOSTAR survey data allows us to observe extended non-thermal emission quite well.  All supernova remnants (SNRs) with previous radio observations present in the catalog of Galactic SNRs \citep{Green2019} were detected in the GLOSTAR data.  Since SNRs are in general linearly polarized, the GLOSTAR polarization data are helpful in identifying new SNRs.

As an illustration, we present  the GLOSTAR Stokes I and the linearly polarized emission ($\sqrt{Q^2 + U^2}$) of the MAGPIS candidate SNR G28.76$-$0.44 \citep{HelfandBecker2006} in Figure \ref{fig:G28.78}.  The linearly polarized flux density was measured in each channel first and then averaged in order to account for bandwidth depolarization effects.  On the linearly polarized flux density map, we also show the electric field vectors corrected for rotation measure (RM).  The RM was measured using a linear fit of polarization angle measured in each channel against $\lambda^2$, where $\lambda$ is the central wavelength of that channel. The mean RM in this region was $-58$~rad~m$^{-2}$ with uncertainties of $25-35$~rad~m$^{-2}$.

We observe that the electric field vectors are preferentially aligned along the shell, implying that the magnetic field is pointing radially outward of the remnant.  A radial magnetic field is typically observed in the shells of young SNRs \citep{Milne1987}.  This is likely due to the effect of magnetic field stretching caused by Rayleigh-Taylor instabilities in the turbulent layer between the ejecta and the ISM \citep{Jun1996}. The detection of polarized emission now finally confirms that the source is indeed a supernova remnant. This object, along with several other candidate SNRs, is discussed in \cite{Dokara2021}.

\begin{figure*}
        \includegraphics[width=0.95\textwidth]{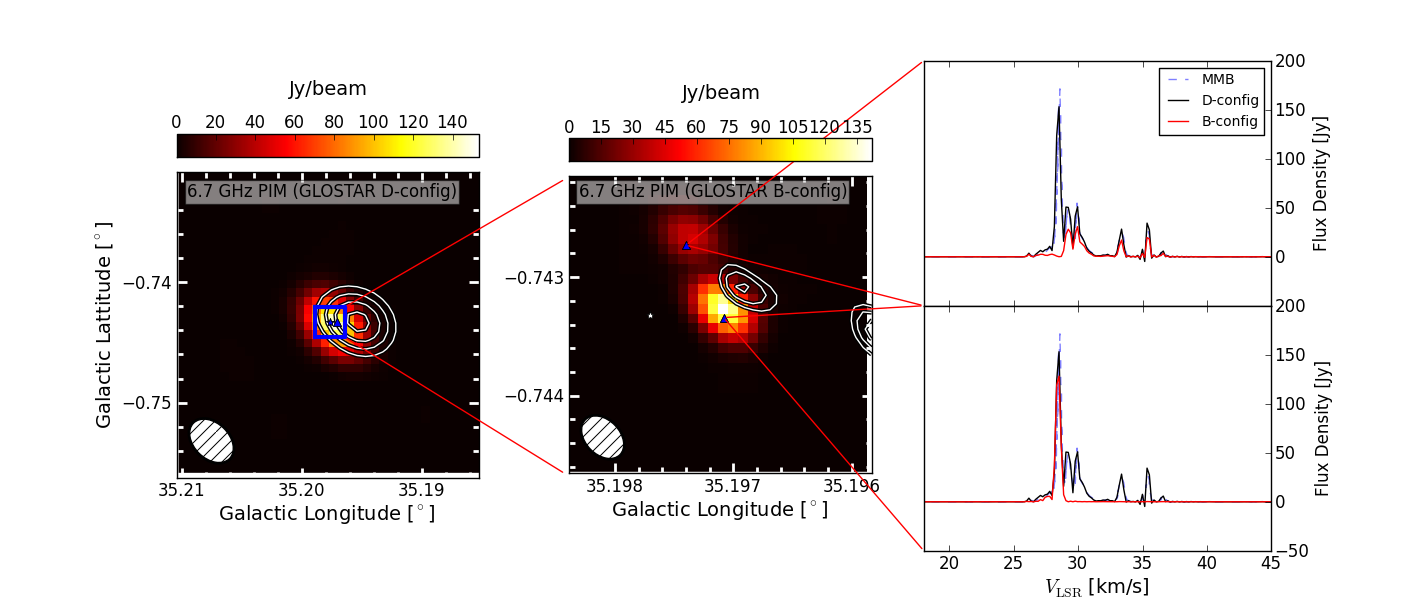}
        \caption{Presented are peak intensity maps (PIM) over the velocity range of the methanol maser source G35.197-0.743 and spectra obtained from VLA data. On the left is the VLA D configuration data, which has an average beamsize of $\sim18$ arcsec. When observed in the B configuration (with beamsize $\sim1.5$ arcsec, middle panel), one can see that the higher resolution allows us to resolve the maser into two sources. Viewing the spectra on the right, we see that the first peak at 28.51\,km s$^{-1}$ belongs only to the southern source while the rest belong to the northern source. The white contours mark the GLOSTAR radio continuum emission. }
        \label{fig:mc_meth}
\end{figure*}

\subsubsection{The hypercompact H{\sc ii} region G28.2-0.04 N}

Hyper-compact H{\sc ii} (HC-H{\sc ii}) regions trace the earliest phases of massive star formation. At the GLOSTAR observed frequencies ($\sim6.0$~GHz) the emission from them is expected to be
optically thick, and with spectral indices of $1-2$ \citep{kurtz2005}. Thus, they are expected to be relatively weak at these frequencies. The sensitivity provided by previous surveys bias the results and made HC H{\sc ii} regions difficult to detect 
\citep{kurtz2005}. The better sensitivity provided by GLOSTAR could partially alleviate this bias. Here, we present an example of a well-known HC H{\sc ii} region and its properties in the GLOSTAR images. 

G28.2-0.04 N, shown in Fig.~\ref{fig:HCHIIG28}, was one of the first radio sources classified as a HC H{\sc ii} region given its properties at radio frequencies \citep[e.g., broad radio recombination lines, rising spectral index, high electron density;][]{sewilo2004}. It has been observed and detected in our GLOSTAR observations. However, the morphology in the final maps is different. 

In the low-resolution image it is detected as a single source and with a total flux of $396\pm20$~mJy \citep[labeled  G028.200-00.050 in our catalog;][]{Medina2019}. On the other hand, in the combined map (VLA D+B) two sources are clearly spatially resolved. These two sources are the components of the system, G028.2-0.04~N and G028.2-0.04~S. The integrated fluxes of these sources are $210\pm4$~mJy and $132\pm4$~mJy. Finally, in the high-resolution map  the northern component is circular, while the southern component is shapted like a shell. Their integrated fluxes are $147\pm7$ and $22\pm1$~mJy.

The in-band spectral index of the northern component is $0.8\pm0.1$, which is consistent within two sigma with earlier values of $1.0\pm0.1$ obtained by \citet{sewilo2004}.
It is worth  highlighting that the earlier estimates 
were obtained with dedicated observations in different bands.

\subsection{6.7\,GHz methanol}

The 6.7\,GHz \meth maser was first discovered by \cite{menten1991} and is  one of the brightest species of masers observed. It has already been shown to be exclusively associated
with young massive stars (\citealt{minier2003}; \citealt{xu2008}), and therefore a sensitive and unbiased survey would trace the early stages of high-mass star formation in our Galaxy.
Furthermore, the distribution of these masers can help us study the structure of the spiral arms within the Milky Way. While previous studies already exist, most notably the Methanol Multibeam (MMB) survey \citep{mmb} and other surveys
(\citealt{pesta2005}; \citealt{Pandian2007}; \citealt{yang2019}), our VLA data is the most sensitive
(RMS$\sim$18\,mJy\,beam$^{-1}$), thus providing potentially the most complete catalog to 
date. In addition to the D configuration data, we also have B-configuration data for a large fraction of the survey in order to provide more accurate positions and observations of multiple epochs. 

G35.197-0.753 (\citealt{xu2008}; \citealt{mmb20_60}) is an example source, and is shown in Fig~\ref{fig:mc_meth}. Plotted are peak intensity maps (PIMs) where we take the highest 
intensity along the velocity range and map it accordingly. From this figure, we can  already see the
benefit the additional higher resolution data provides as we are able to clearly differentiate
two sources spatially. Inspecting the spectra, we  also find that the brightest component seen in the D-configuration data at 28.51\,km s$^{-1}$ originates in the southern source, while the
remaining velocity components belong only to the northern source.

\begin{figure}
        \includegraphics[width=0.5\textwidth]{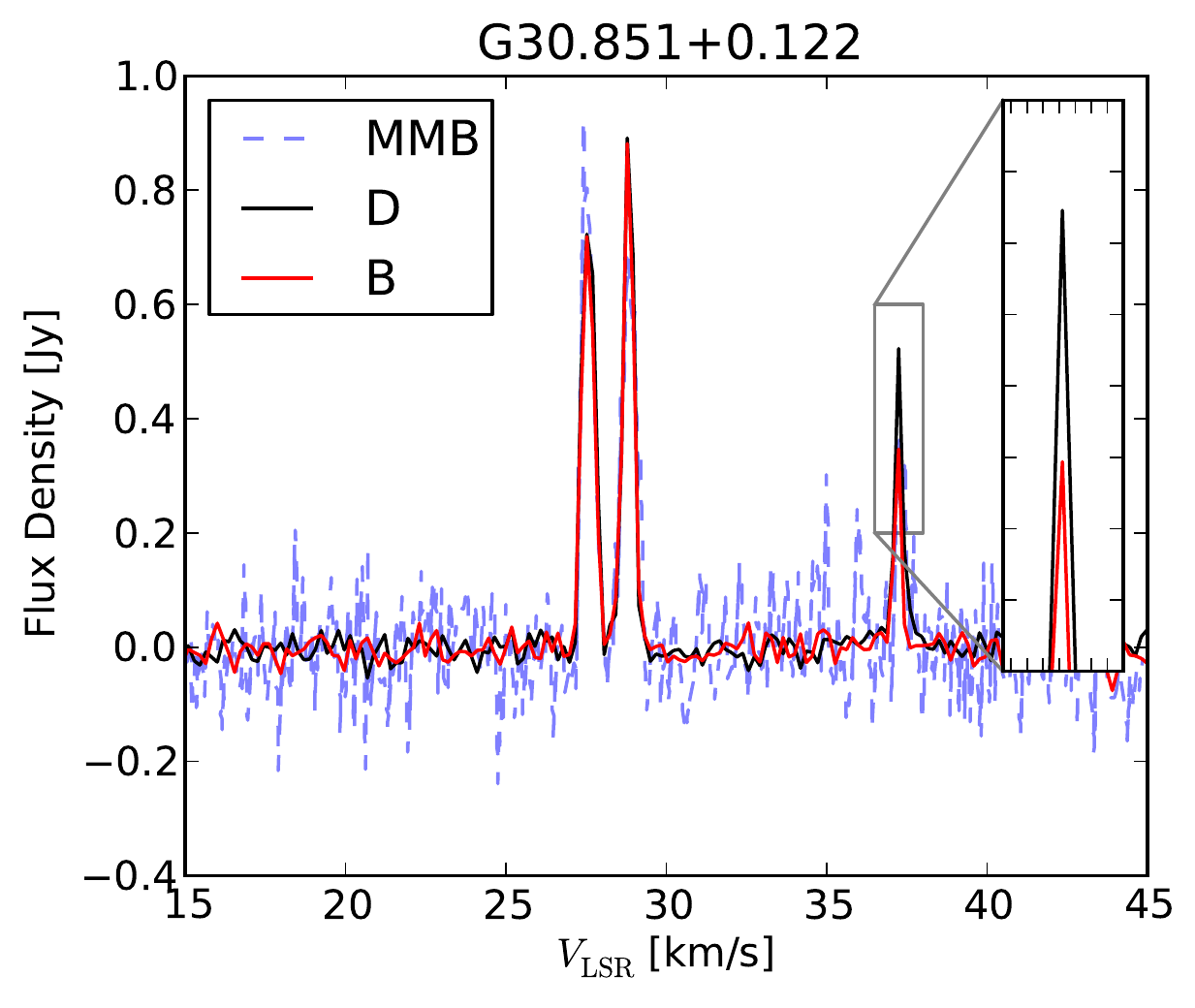}
        \caption{Methanol maser spectra of source G30.851+0.122. Shown   are the GLOSTAR spectra in black and red solid lines for D configuration and B configuration, respectively. The dashed blue line is the MMB spectrum of the corresponding source. The right inset depicts a zoom-in of the  velocity component, which shows a difference in intensity between the two configurations; it is spatially unresolved in  D and in B configuration.}
        \label{fig:var_meth}
\end{figure}

Furthermore, our access to multiple epochs allows us to observe potential maser variability that is not yet well understood. Maser variability typically occurs with periods of $\sim$20 to 200 days and with sinusoidal patterns or intermittently with a quiescent phase (\citealt{goedhart2004}; \citealt{szymczak2018}; \citealt{olech2019}).
There are two kinds of models to explain the observed variability. The first explains the variation through cyclic variations in the seed photon flux from a colliding-wind binary (CWB; \citealt{vanderwalt2009}, \citealt{vanderwalt2011}). The second type of model assumes changes in the temperature of the dust grains in the environment of the masing region. This affects the infrared radiation flux, and thus the pump rate. There are multiple hypotheses for the origins of the changes in the temperature, such as periodic accretion from a circumbinary accretion disk \citep{araya2010},
stellar pulsation (\citealt{inayoshi2013}; \citealt{sanna2015}), and spiral shock waves in the disk of a binary system (\citealt{ps2014}; \citealt{szymczak2018}).

We show in Fig.~\ref{fig:var_meth} the
peak flux density of G30.851+0.122  of the two configurations and  the MMB
spectrum. Here we see that our sensitivity appears to be better by a factor of $\sim$10 compared to the MMB. We also see that in the GLOSTAR data, the first two velocity components match, while
the third component decreases to roughly 60\% over the course of 8 months (the time between
observations of VLA D configuration and B configuration). The combination with other known surveys with GLOSTAR can give us an even broader glimpse into the nature of methanol maser variability.

\subsection{4.8\,GHz formaldehyde}
It has been known for some time  that \form is associated with massive star-forming regions. Furthermore, \cite{downes1980} found that 80\% of the H{\sc ii} regions
surveyed had \form absorption. In particular, the $1_{10}-1_{11}$ line of \form is mainly observed in
absorption at 4.8\,GHz and is associated with interstellar dust clouds against the 2.73\,K
cosmic microwave background (CMB) \citep{snyder1969}. It has been found that these absorption 
features can be used to trace the kinematics of infrared dark clouds (IRDCs) and are associated 
with infrared sources \citep{okoh2014}. Furthermore, \form is ubiquitous and is 
therefore a good probe of the physical conditions of molecular clouds with densities 
$10^3-10^5$ cm$^{-3}$ \citep{mangum1993}, and thus the cold and dense molecular clouds.

We follow a procedure similar to our methanol analysis.
The analysis of the pilot region  resulted in a detection of 20 compact \form absorption sources. 
Compared with the catalogs of \cite{downes1980} and \cite{okoh2014}, this corresponds to
a detection rate of $\sim40\%$. The strongest sources in their catalogs were all detected
automatically from our source extraction, while three other  sources were found after a manual inspection.
The discrepancy can be explained from the much longer integration time used by previous
surveys ($>$15\,minutes) that would allow them to detect many more weaker sources. Furthermore, their observations have resolutions
of 2.6\arcmin\ and 10\arcmin, respectively, which is a factor of 5--20 higher than ours. It is then plausible that some of these sources were resolved out due to our much higher 
angular resolution.

\begin{figure}
        \includegraphics[width=0.5\textwidth]{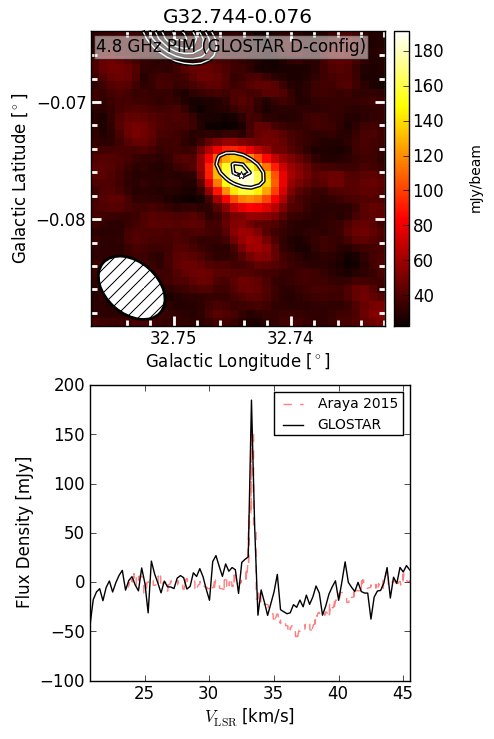}
        \caption{Formaldehyde maser in G32.744-0.076. \textit{Top:} Peak intensity map of known 4.8\,GHz formaldehyde maser G32.744-0.076 \citep{araya2015} using GLOSTAR VLA D-configuration data and the continuum emission (white contours). \textit{Bottom:} Spectra extracted from GLOSTAR (solid black line) and  from the \cite{araya2015} paper (dashed red line). Despite the lower sensitivity of our data,  the  faint superimposed absorption feature can be seen.}
        \label{fig:form_maser}
\end{figure}

First detected by \cite{downes1974}, formaldehyde was also found to show maser emission at 4.8\,GHz.
However, despite dedicated attempts to detect these molecules again, formalehyde remains very rare and seem to
be exclusively
related to the high-mass star formation process. Within the pilot region, only the G32.744-0.076 \form maser \citep{araya2015} was detected, and is shown in Fig.~\ref{fig:form_maser}, while the maser G29.96-0.02 reported by \cite{pratap1994} was not detected.

\begin{figure*}
\includegraphics[width=19cm]{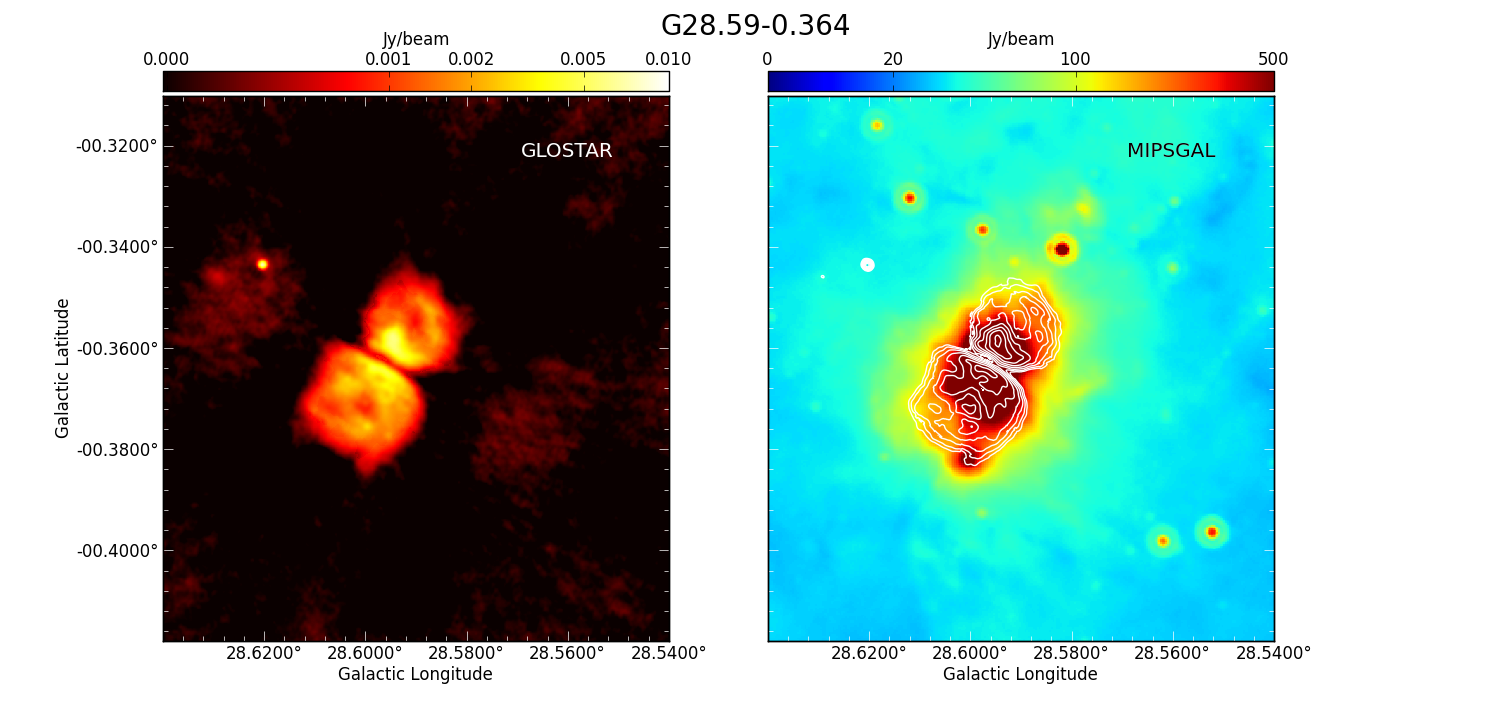}
\caption{Continuum image of the H{\sc ii} region G28.594-0.359 from the combined VLA D+B-configuration data (left), and an MIPSGAL 24 $\mu$m image with the radio contours overlaid (right). The diffuse emission on both sides are most likely weak residual side lobes.}
\label{fig:G28.59}
\end{figure*}

\subsection{Radio recombination lines}
Radio recombination lines are tracers of ionized gas in \ion{H}{ii} regions and have been used to confirm \ion{H}{ii} region candidates detected in near-infrared emission \citep[e.g.,][]{AndersonBania:2014aa}. They provide kinematic and therefore distance information, and allow us to study of physical properties of the emitting gas, such as the determination of the electron temperature \citep[e.g.,][]{QuirezaRood:2006aa} and the feedback of \ion{H}{ii} regions on the surrounding molecular cloud environment.
We highlight two examples of RRL emission detected with the GLOSTAR survey for its pilot region. The $\sim$100 RRL detections in the pilot region will be described in greater detail in forthcoming works. Here, we show an example of a relatively isolated region of star formation, G28.694-0.359, and possible interaction with the surrounding molecular gas. 
We show the different aspects that the GLOSTAR data tell us about high-mass star formation via the example of the massive star-forming complex G29.93-0.03.

\subsubsection{H{\sc ii} region G28.594-0.359}

\begin{figure*}
  \includegraphics[width=0.9\textwidth]{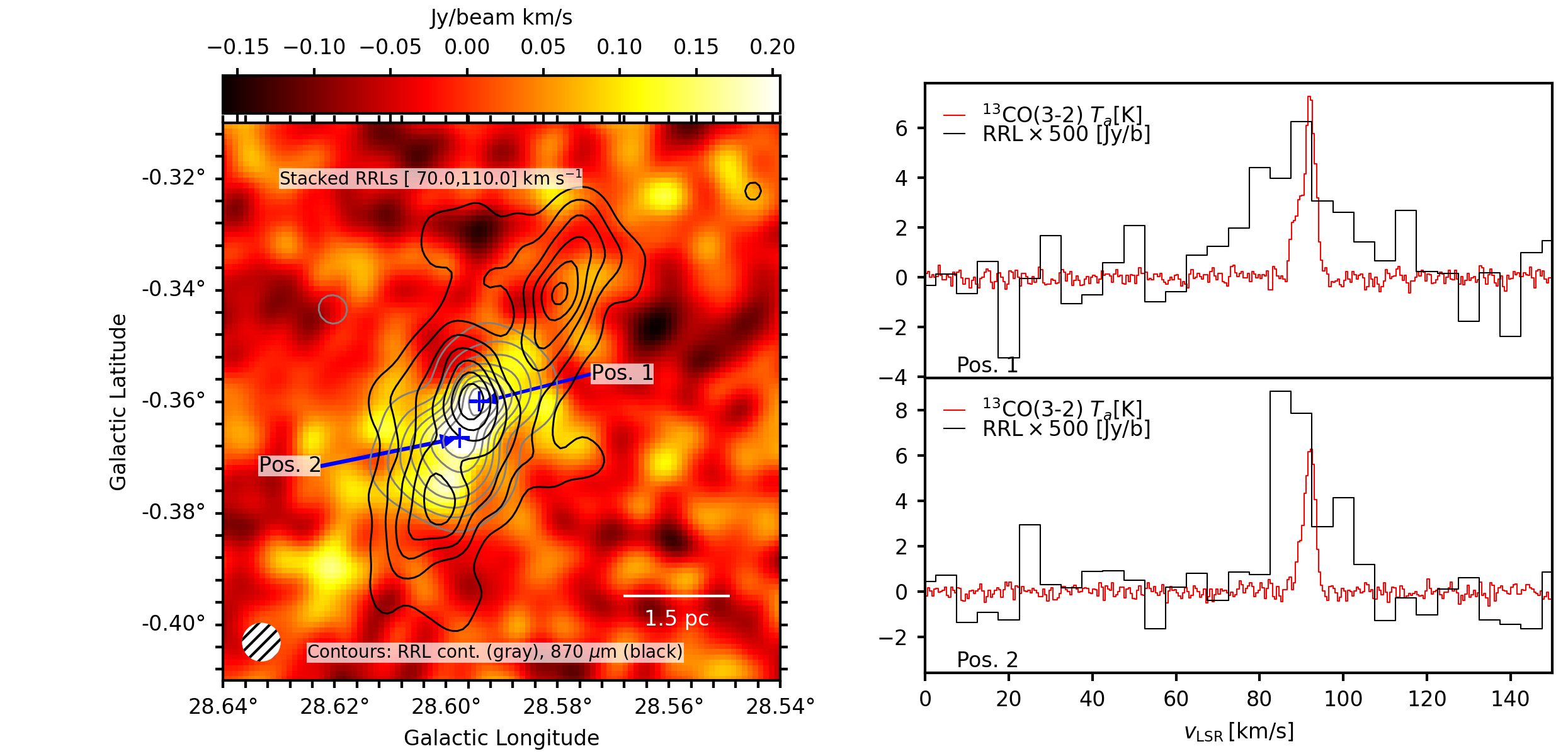}
\caption{Overview of the RRL emission of the \ion{H}{ii} region G28.59-0.364. The left panel shows integrated, stacked RRL emission at 25\arcsec\ resolution. The GLOSTAR broadband 5.8 GHz continuum (gray contours) from $-$0.005\,\jyb in steps of 0.01\,\jyb. ATLASGAL $870\,\mu$m emission \citep{SchullerMenten:2009aa} is shown (black contours) starting from 0.2\,\jyb in steps of 0.2\,\jyb. The right panel shows \mbox{${}^{13}$CO (3-2)} emission (in red) from the CHIMPS survey \citep{RigbyMoore:2016aa}. All data have been smoothed to the angular resolution of the RRL data of 25\arcsec. The positions of extracted spectra in the side panels are indicated with blue crosses.}
  \label{fig:overview_g28}
\end{figure*}

The \ion{H}{ii} region G28.594-0.359 shows two lobes that are resolved in the GLOSTAR B+D configuration continuum data (Fig.~\ref{fig:G28.59}). The RRLs indicate a LSR velocity of 90 \kms. Using the parallax-based distance calculator from \cite{ReidDame2016}, we get a distance of 4.5 $\pm$ 0.4\,kpc. The ATLASGAL 870 $\mu$m sources associated with the \ion{H}{ii} region show similar distances of 4.7--4.8\,kpc \citep{UrquhartKonig:2018aa}. The lobes have a radius of 0.5-0.9\,pc. The strongest emission of both lobes occurs towards the center of the complex. In the region between the two bubbles, the continuum emission seems to be strongly reduced. At the same time, as seen in Fig.~\ref{fig:overview_g28}, dust continuum emission at 870\,$\mu$m appears to lie in between the two bubbles. This constellation resembles the famous bipolar \ion{H}{ii} region S106 \citep[e.g.,][]{BallySnell:1983aa}, and it is likely that both lobes share a common origin within the dust ridge separating the two lobes. In contrast to S106, no ionizing source is detected in the 5.8\,GHz data between the lobes.

\begin{figure}
\includegraphics[trim=80 0 80 0,clip,width=1.0\columnwidth]{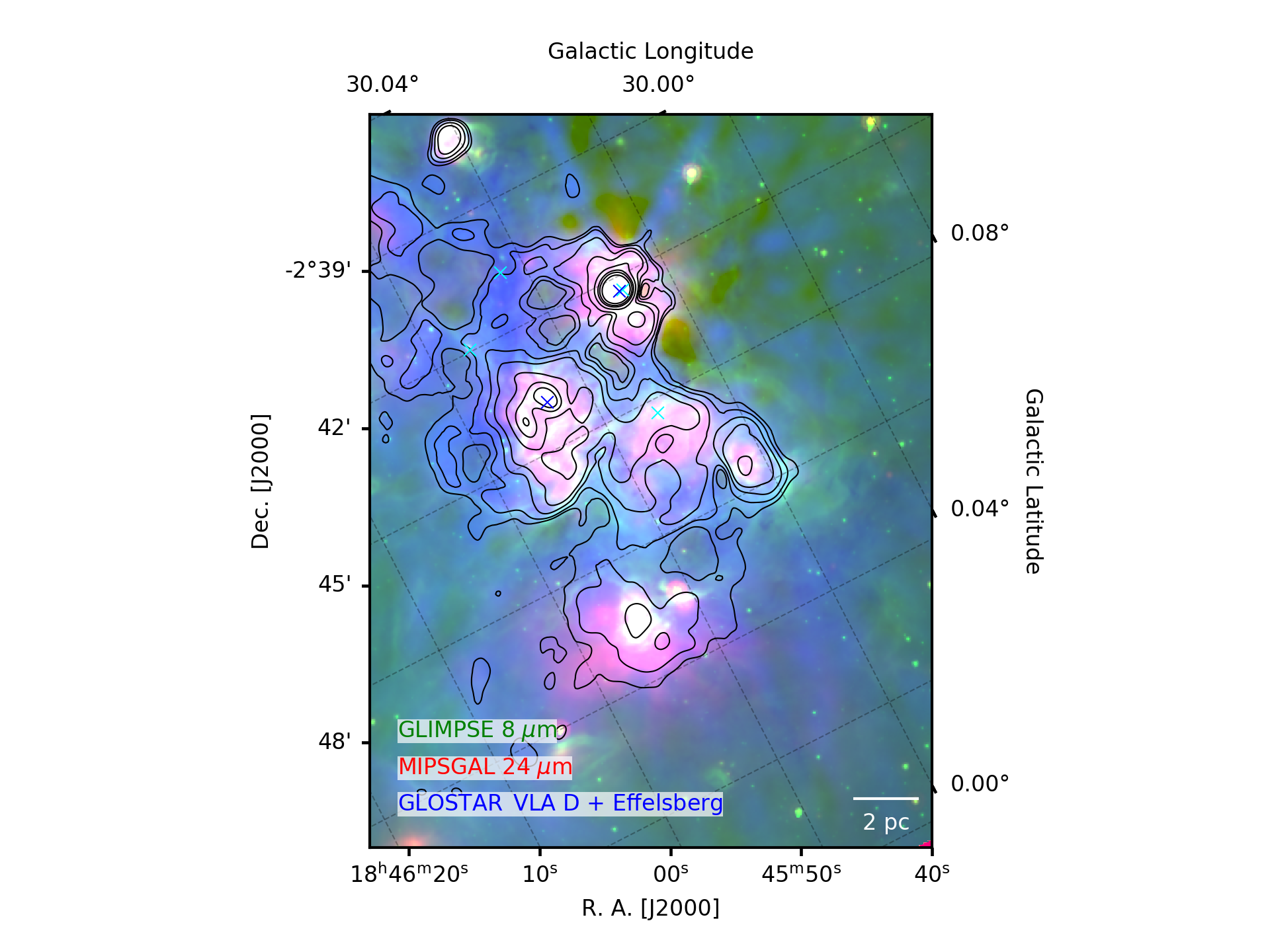}
\caption{Multi-color image of GLIMPSE 8 $\mu$m \citep[{\it green},][]{churchwell2009}, MIPSGAL 24 $\mu$m \citep[{\it red},][]{carey2009},  and GLOSTAR VLA D+Effelsberg 5.8 GHz emission ({\it blue}) of the G29.93-0.03 (W43-south) star formation complex. Square-root scaling was used and  the individual data sets were truncated at 500 and 1000 MJy/sr for the 8 and 24 $\mu$m data, respectively, and at 20 m\jyb for the centimeter emission. Saturated pixels in the MIPSGAL maps were replaced with the truncation value. The data are shown in their native resolution. For reference,  the markers from Fig.~\ref{fig:overview_g29} are overplotted. The GLOSTAR 5.8 GHz data are repeated as contours for clarity, in steps of 15, 20, 30, 45, 100, 200, 300, and 500 m\jyb.}
\label{fig:overview_g29_rgb}
\end{figure}

\begin{figure*}
\centering
\includegraphics[width=1.0\textwidth]{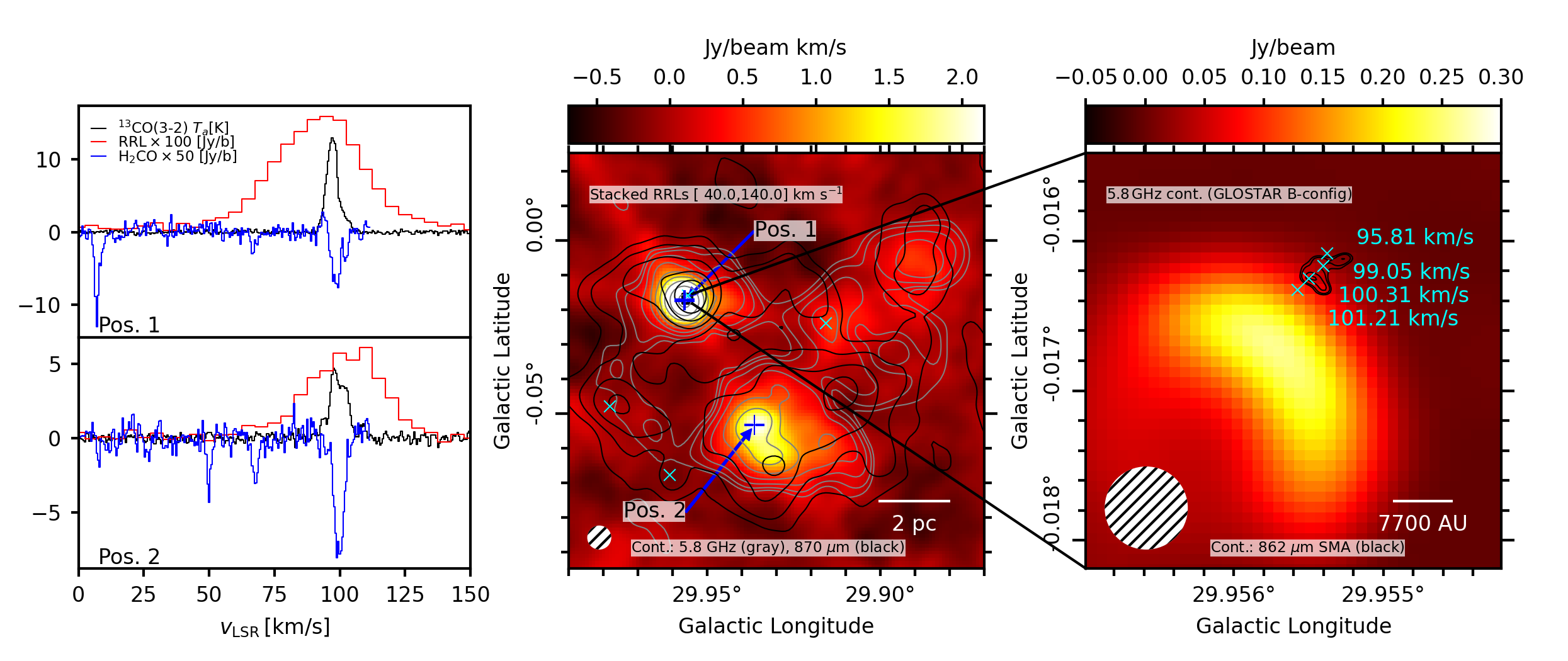}
\caption{Overview of the RRL emission in the star-forming complex around $l=29.93^\circ$, $b=-0.03^\circ$. The central panel shows integrated emission between 40 and 140 \kms\ after stacking seven RRLs in velocity at 25\arcsec\ resolution. The GLOSTAR 5.8 GHz D-array continuum is shown (gray contours) always increasing by a factor of two starting from 10\,m\jyb. ATLASGAL $870\,\mu$m emission \citep{SchullerMenten:2009aa} is shown (black contours) increasing by a factor of two starting from 0.5\,\jyb. The positions of extracted spectra in the side panels are indicated with blue crosses; detected CH$_3$OH 6.8\,GHz masers are indicated by cyan crosses. The side panels on the left compare the RRL spectra to $\mathrm{^{13}CO(3-2)}$ emission from the CHIMPS survey \citep{RigbyMoore:2016aa}, and to the ${\rm H_2CO}$ absorption at 0.5\,\kms\ spectral resolution from GLOSTAR. All data have been smoothed to 25\arcsec\ resolution. The right panel shows a zoom-in on the cometary \ion{H}{ii} region G29.96-0.02 in 5.8 GHz continuum emission from the GLOSTAR survey in B configuration at 2\arcsec\ resolution. Overlaid with cyan crosses are fitted ${\rm CH_3OH}$ maser position peaks at different velocity channels from GLOSTAR (1.4\arcsec$\times$1.0\arcsec). SMA 862\,$\mu$m emission from \citet{BeutherZhang:2007aa} is shown (black contours) at 83, 105, 147, and 168\,m\jyb.}
\label{fig:overview_g29}
\end{figure*}

The lobes blur into two barely resolved emission peaks at 25\arcsec\ resolution. We detect RRL emission towards both peaks, and fit both spectra assuming Gaussian line profiles. The emission peak is weakly detected at 13$\pm$3\,m\jyb\ and 18$\pm$3\,m\jyb\ for posistions~1 and 2, respectively. While this corresponds to only 4-6$\sigma$, the transition is detected in multiple channels for which we are confident of a detection. 
As  seen in the right panels of Fig.~\ref{fig:overview_g28}, the velocity of the RRL emission roughly agrees with the peak velocities of the \mbox{${}^{13}$CO (1-0)} emission across the region, implying that molecular and ionized gas are kinematically connected. We note that small differences exist. Towards positions~1 and 2, the RRL velocities appear slightly blueshifted from the main peak of the CO emission, which is even more clearly seen at higher angular resolution in the \mbox{${}^{13}$CO (3-2)} transition obtained from the CHIMPS survey \citep{RigbyMoore:2016aa}. This may be indicative of expansion of gas away from the host molecular cloud.

Morphological, the location of ionized and molecular gas may indicate a physical connection between them. As an indicator for the morphology of the dense molecular gas we use the 870\,$\mu$m dust emission (Fig.~\ref{fig:overview_g28}, black contours) from ATLASGAL \citep{SchullerMenten:2009aa} as well as the integrated ${}^{13}$CO (3-2) emission between 86\,\kms\ and 96\,\kms \citep{RigbyMoore:2016aa}. Three ATLASGAL clumps \citep{ContrerasSchuller:2013aa,UrquhartCsengeriWyrowski2014} are located in the region. The source closest to the position of the continuum emission, AGAL028.596-00.361, overlaps with the position of the gap between the two  radio lobes, and peaks slightly offset towards higher Galactic latitudes. The other two sources are located below the radio lobe at low Galactic latitudes, AGAL028.601-00.377, and above the radio lobe at higher Galactic latitudes, AGAL028.579-00.341. All three ATLASGAL sources are classified as young stellar objects in \citet{UrquhartKonig:2018aa}, and have similar masses of $\log M/M_\odot = 2.9$ and luminosities of $\log L/L_\odot$= 4.4, 4.1, and 3.8, respectively. The sources appear to be morphologically connected to the radio lobes, as the 5 cm continuum emission extends towards both sources and also slightly brightens towards them (Fig.~\ref{fig:overview_g28}; the same holds for the emission at 24\,$\mu$m). We speculate that the \ion{H}{ii} regions have either cleared out the gaps seen in the 870\,$\mu$m emission in an originally elongated molecular cloud during their expansion, or that the radio lobes simply expanded into spaces of less dense molecular gas, possibly originating from fragmentation of the parent molecular cloud into three denser clumps.

\subsubsection{G29.93-0.03}
We chose the \ion{H}{ii} region and molecular cloud complex G29.93-0.03, also known as W43-south, as a demonstration region for the collective use of the different observed tracers in the GLOSTAR survey.  W43-south is a site of active star formation \citep[e.g.,][]{ReifensteinWilson:1970aa,SmithBiermann:1978aa}. Together with the high-mass star-forming complex W43-main it constitutes the W43 star-forming complex. Both clouds are potentially physically connected \citep{Nguyen-LuongMotte:2011aa}. The entire complex is located at an average distance of 5.5\,kpc \citep{ZhangMoscadelli:2014aa} in the Scutum-Centaurus arm and at the end of the Galactic bar \citep{Nguyen-LuongMotte:2011aa,CarlhoffNguyen-Luong:2013aa}. Their interaction may give rise to the complex dynamics of the region \citep[e.g.,][]{Nguyen-LuongMotte:2011aa,Beuther2017}. Since the properties of the ionized gas in this region are well known \citep[e.g.,][]{BeltranOlmi:2013aa}, W43-south is well-suited to demonstrate the survey data.

With GLOSTAR, we observe tracers of different stages of the star formation process, and at the same time we observe these at different scales. With Effelsberg we provide important short-spacing information for the interferometry maps. In terms of spatial scales, the Effelsberg observations trace extended continuum radiation at C band   from thermal and non-thermal emission at a spatial resolution of 3-4 pc at the distance of W43-south, while the VLA observations allow us to map the \ion{H}{ii} region complex at a resolution of $\sim$0.5 pc (D-configuration data) and to distinguish individual (ultra-)compact \ion{H}{ii} regions at scales of $\sim$0.05 pc or $\sim$10000 AU (B-configuration data). Similar considerations hold for ${\rm H_2CO}$ and the RRLs, for which the observations give information   on the extended and on the small-scale structures.

W43-south  hosts many \ion{H}{ii} regions, powered by multiple O and B stars \citep[e.g.,][]{BeltranOlmi:2013aa}. Figure~\ref{fig:overview_g29_rgb} shows a three-color composite image of Spitzer IRAC 8 $\mu$m, MIPSGAL 24 $\mu$m, and GLOSTAR VLA D+Effelsberg 5.8 GHz emission. The region harbors multiple \ion{H}{ii} regions. Some of them have formed cavities in the surrounding molecular cloud, which are signs of already more advanced stages of star formation. They are filled with 24 $\mu$m emission, tracing hot dust, and 5.8 GHz emission from free-free emission. The cavities are confined by photon dominated regions (PDRs) that can be most clearly seen in the high-resolution 8 $\mu$m data, tracing polycyclic aromatic hydrocarbon (PAH) emission.

The well-studied hot core G29.96-0.02 is an example of high-mass star formation in an early stage \citep[e.g.,][]{CesaroniChurchwell:1994aa,CesaroniHofner:1998aa,GibbWyrowski:2004aa}. In close proximity is a well-known cometary \ion{H}{ii} region \citep[e.g.,][]{WoodChurchwell:1989ab}, which can be seen in the GLOSTAR B-configuration data in Fig.~\ref{fig:overview_g29}. Starting from the inlay on the right, the image shows the positions of different methanol maser emission peaks determined from the GLOSTAR B configuration towards the hot core G29.96-0.02 \citep[e.g.,][]{WalshBurton:1998aa,MinierConway:2001aa}. We fitted the positions of the peaks at different velocities with 2D Gaussians and find the velocity to be decreasing with distance to the UC\ion{H}{ii} region. While the precise overall location of the masers with respect to the dense molecular gas may slightly be affected by systematic uncertainties (see Sect.~\ref{sect:astrometric_accuracy}), the direction of the velocity gradient roughly follows the known gradients seen in molecular gas tracers \citep[e.g.,][]{OlmiCesaroni:2003aa,BeutherZhang:2007aa,BeltranCesaroni:2011aa}. It also appears to be misaligned with the direction of the known molecular outflow of the source \citep{GibbWyrowski:2004aa}.

Turning to the entire \ion{H}{ii} region complex, the middle panel of Fig.~\ref{fig:overview_g29} shows the integrated RRL emission across the region and methanol masers recovered in the GLOSTAR data. Further spectral lines from the GLOSTAR survey can be seen in spectra shown in the left panels of  Fig.~\ref{fig:overview_g29}. The spectra are shown towards position~1, the cometary \ion{H}{ii} region at G29.96-0.02 and towards position~2 which is a bright \ion{H}{ii} region, whose ionized gas radiation is equivalent to be powered by a star of spectral type O5, the most powerful star in the region \citep{BeltranOlmi:2013aa}. We map the widely studied formaldehyde transition at 4.8\,GHz (blue; e.g., \citealt{Wilson:1972aa}) and find it to be well-matched to other molecular gas tracers, such as the \mbox{${}^{13}$CO(3-2)} emission shown here. The stacked RRL emission spectra shown in red are two examples of the widespread RRL emission seen in the middle panel of Fig.~\ref{fig:overview_g29}, which provide kinematic information on the ionized gas. This brief overview of W43-south hence illustrates the quality of the data and possibilities for future study.

\section{Conclusion}
We presented an overview of a new radio continuum and spectral line survey of 145 square degrees of the northern Galactic plane obtained with the VLA and the 100-m Effelsberg telescopes in the frequency range of 4--8 GHz. After a detailed description of the observational setup, and the calibration and imaging techniques we presented a few selected scientific results that show the quality of the survey data and allows a glimpse of the broad range of scientific topics that can be addressed by the GLOSTAR survey. These will be investigated in more detail in several upcoming publications that are already in preparation.

\begin{acknowledgements}
We thank the anonymous referee for providing useful suggestions and comments, which helped to improve the manuscript. This research made use of Astropy,\footnote{http://www.astropy.org} a community-developed core Python package for Astronomy \citep{Astropy-CollaborationRobitaille:2013ab,Astropy-CollaborationPrice-Whelan:2018aa}, and related packages such as Spectral-Cube and Reproject, as well as of APLpy \citep{RobitailleBressert:2012aa}, an open-source plotting package for Python. Early work on the project described here was supported by the European Research Council (ERC) via the ERC Advanced Investigator Grant 247078 (\textit{GLOSTAR: A Global View of Star Formation in the Milky Way}) awarded to K. M. M, and a Marie Curie International Outgoing Fellowship (No. 275596) awarded to A.B., while H. B. acknowledges support from the Horizon 2020 Framework Programme via the ERC Consolidator Grant CSF-648505. H.B. also acknowledges support from the Deutsche Forschungsgemeinschaft in the Collaborative Research Center (SFB 881) “The Milky Way System” (subproject B1).
N.R. acknowledges support from MPG through Max-Planck India partner group grant.
\end{acknowledgements}

\bibliographystyle{aa} 
\bibliography{brunthaler_refs} 
\end{document}